\documentclass[prd,onecolumn,nofootinbib,notitlepage]{revtex4-1}
\usepackage{amsfonts,amsmath,amscd,amssymb}
\usepackage{graphicx}
\usepackage{bbm}
\usepackage[utf8x]{inputenc}
\usepackage[unicode=true,
 bookmarks=false,
 breaklinks=false,pdfborder={0 0 1},backref=section,colorlinks=false]
 {hyperref}
\usepackage{ulem}

\newcommand{\cobicross}{{\triangleright\!\!\!\blacktriangleleft}}

\begin{document}

\title{Planck-scale-deformed relativistic symmetries and diffeomorphisms in momentum space}

\author{Giovanni AMELINO-CAMELIA}
\affiliation{Dipartimento di Fisica Ettore Pancini, Università di Napoli ”Federico II”, and INFN, Sezione di Napoli, Complesso Univ. Monte S. Angelo, I-80126 Napoli, Italy}

\author{Stefano BIANCO}
\affiliation{Max Planck Institute for Gravitational Physics (Albert Einstein Institute),\\
 Am Mühlenberg 1, 14476 Potsdam-Golm, Germany}

\author{Giacomo ROSATI}
\affiliation{Institute for Theoretical Physics, University of Wroc\l{}aw, Pl.\ Maksa Borna 9, Pl--50-204 Wroc\l{}aw, Poland }

\begin{abstract}
We study the implications of a change of coordinatization of momentum space for theories with curved momentum space.
We of course find that after a passive diffeomorphism the theory yields the same physical predictions, as one would expect considering that a simple reparametrization should not change physics. However, it appears that general momentum-space covariance (invariance under active diffeomorphisms of momentum space) cannot be enforced, and within a given set of prescriptions on how the theory should encode momentum-space metric and affine connection the physical predictions do depend on the momentum space background. These conclusions find support in some
general arguments and in our quantitative analysis of a much-studied toy model with maximally-symmetric (curved) momentum space.
\end{abstract}

\maketitle

\section{Introduction}
\label{sec:Intro}

Several recent studies have been devoted to the possibility of Planck-scale-deformed relativistic symmetries~\cite{DSR,kowadsr,leedsrPRL},
and particularly to the case in which the deformation is due to the presence of curvature on
momentum space~\cite{JurekDeSitt,FreLiv3Dk,JurekFrekfield1e2,ArzanoQfieldCurved,
RelLocPrinciple,JurekReview,CarmonaRelLoc,anatomy,FlaGiukRelLoc,
FreidelFieldCurved,GirelliLivineSnyder,FreidelSnyderRelLoc,MigSamSnyderRL}. This leads to scenarios which are conceptually intriguing and
often provide opportunities for quantum gravity phenomenology~\cite{QGphen}.
The idea of a curved momentum space, with the scale of curvature driven by Planck energy $E_P \sim 10^{-19}\text{GeV}/c^2$, can be dated back to the seminal work of Max Born~\cite{BornReciprocity}.
A notion of duality between curvature of momentum space and  spacetime non-commutativity was first discussed in studies by Snyder~\cite{snyder}.
These ideas find a particularly satisfactory formalization in the context of Hopf-algebras~\cite{MajidFoundation}, a mathematical framework suitable for describing on the same footing non-commutative spacetimes and their associated symmetries, so that the curvature of momentum space emerges naturally as the dual aspect of spacetime non-commutativity~(see for instance \cite{JurekFrekfield1e2}).

The ``relative-locality framework''~\cite{RelLocPrinciple} has been adopted in several studies as the basis for a Lagrangian description of the relativistic kinematics of particles in theories with a curved momentum space.
The label ``relative locality'' reflects the fact that
in theories with deformed relativistic symmetries (also known as DSR-relativistic theories~\cite{DSR,kowadsr,leedsrPRL}),
due to
the presence of a second invariant scale $\ell$ with dimensions of inverse-momentum,  one typically finds that the standard absolute notion of spacetime locality must be replaced by a relative one~\cite{whataboutbob,kappabob} (in close analogy to how the adoption of the speed-of-light relativistic invariant leads to the replacement of absolute simultaneity with relative simultaneity).

The Lagrangian formulation prescribed by the relative-locality framework~\cite{RelLocPrinciple} gives the on-shell relation
in terms of the metric on momentum space, and free-particle propagation is described in manifestly-covariant manner as a evolution in an
affine parameter governed by the on-shell-relation Hamiltonian. The affine connection is then used to formulate suitable boundary terms for particle worldlines ending at an interaction vertex, thereby specifying the form  of energy-momentum conservation
and of the generators of spacetime translations.
The picture will be relativistic (but in general with deformed, DSR-relativistic, laws) if certain conditions are satisfied~\cite{GACboosts,CarmonaBeySpecRel}, as it happens for maximally-symmetric momentum spaces~\cite{palmisano,CarmonaRelancioComposition}.
Several examples of these relativistic scenarios have been considered in recent years~\cite{CarmonaRelLoc,anatomy,FlaGiukRelLoc,spinning,FreidelSnyderRelLoc,MigSamSnyderRL,causality,palmisano,MigRosSnyderRelLoc,multipart,CarmonaLocal,GiuliaBoosts,CarmonaRelancioComposition}.

The aim of this paper is to study the properties of such a relativistic theory with curved momentum space under changes of the
coordinatization of momentum space.
This has been so far only contemplated by assuming that momentum-space general covariance could be adopted axiomatically~\cite{FreSmoGRBrelLoc,FreidelSnyderRelLoc,meljanac}, essentially assuming that curvature of momentum space should
 be accompanied by general covariance just on the basis of an analogy with how spacetime curvature is
 in some sense (within general relativity) associated with general covariance; however, as we shall here show, once the theory is fully specified
its properties under changes of the
coordinatization of momentum space are fixed (and therefore could not possibly be imposed externally by axiomatization).

Crucial for our analysis is the difference between invariance under passive diffeomorphisms and invariance under active diffeomorphisms.
A passive-diffeomorphism transformation just applies the change of coordinatization of momentum space to all aspects of the theory,
including the observable aspects, and the physical content of the theory is automatically invariant under such passive diffeomorphisms,
in the sense that each physical prediction is trivially mapped into the corresponding prediction via the action of the diffeomorphism
transformation. The presence of this sort of invariance under passive diffeomorphisms of course does not ensure momentum-space
background independence.
Background independence requires invariance under active diffeormorphism, whose presence is signaled~\cite{rovelliBook} by the
invariance in form of the action (the equations
of motion) under a change of momentum-space coordinatization, provided that this
is achieved without introducing any external non-dynamical tensor\footnote{In a theory which is background dependent one  may still find a reformulation of the theory, obtained by the ad hoc introduction of an external
non-dynamical tensor, such that the form of the action is formally invariant
under changes of coordination, but in that case the background dependence is encoded in the properties of the external tensor~\cite{rovelliBook}. }.

Following Ref.~\cite{RelLocPrinciple} we assume that the on-shell relation is obtained from the geometry of momentum space by
computing the geodesic distance from the origin to a generic point of momentum space. We shall show that
the on-shell relation changes its form under changes of momentum-space coordinatization, though this works out in just such a way
that the geodesic distance is invariant.
Still following Ref.~\cite{RelLocPrinciple} our actions include some  boundary terms describing conservation laws
at particle interactions, and these boundary terms are specified by the affine connection on momentum space.
By studying the behaviour of these boundary terms under changes of coordinatization of momentum space
we expose a clear mechanism by which the theories are not invariant  under active diffeomorphisms, i.e. their physical predictions  are not independent of the ``momentum space background''.

For most of our analysis we rely on a specific  relativistic framework with curved momentum space, the so
called ``$\kappa$-momentum space''~\cite{JurekDeSitt,JurekFrekfield1e2,anatomy,FlaGiukRelLoc,causality,multipart,CarmonaLocal, GiuliaBoosts}.
This is the curved-momentum-space scenario associated to $\kappa$-Minkowski non-commutative spacetime and its dual,
the $\kappa$-Poincar\'e Hopf-algebra~\cite{Lukierski,MajidRuegg,gacMajid}. Notably $\kappa$-momentum space
 is the group manifold $AN_3$, coinciding with (half of) de Sitter space~\cite{JurekDeSitt,JurekFrekfield1e2,anatomy,FlaGiukRelLoc,causality,multipart,CarmonaLocal}.

In the following we use units for which the velocity of light, $c$, is set to 1, and we assume the deformation scale to be positive, $\ell>0$.

\section{Diffeomorphisms in curved momentum space: the relative locality framework}

In this section we consider the effect of a change of coordinates (passive diffeomorphisms) in a (relativistic) curved momentum space within the relative locality framework~\cite{RelLocPrinciple}.
We start by characterizing the main features of the formalism, outlining the construction of the relative locality action.
We then consider the effect of a change of coordinates on the two main components of the action: the bulk, consisting in the particle on-shell relations, and the boundary terms, characterizing interactions.
We first study the problem from a general perspective, postponing a specific example, based on a specific choice of momentum space, to the following section.

\subsection{The relative locality framework}
\label{sec:RelLoc}

In the relative-locality framework~\cite{RelLocPrinciple}, the deformed relativistic kinematics of a system of interacting particles is encoded in the geometric properties of momentum space.
Considering the momentum space to be described by a (generally curved) manifold ${\cal P}$, one takes a coordinate system $p_\mu$ on ${\cal P}$.
Taking a point $P$ in ${\cal P}$, the geodesic distance $D(0,p_{\mu})$ from the origin of the coordinate system to the point $p_\mu \equiv p_\mu(P)$ describes an orbit in momentum space, as the set of points (a curve in momentum space) for which the geodesic distance has the same value.
The value of the geodesic distance is interpreted as the mass of a particle of momentum $p_\mu$, and it takes a specific expression in terms of coordinates $p_\mu$ that is interpreted as the on-shell relation for the particle, which therefore is encoded in the metric $g^{\mu\nu}$ on the momentum space.
Of course the existence of the orbit depends on the topological properties of ${\cal P}$, and thus on the symmetries of momentum space.
From the relativistic perspective we are interested in cases in which the curvature of momentum space still allows the description of spacetime symmetries (in the generalized DSR-deformed sense). This restricts the choice of possible momentum space metrics.
In particular if one wants to still have a full set of (DSR-) relativistic symmetries, corresponding to 1 time and 3 space translations, 3 rotations, and 3 boosts, the choice of geometries is restricted to maximally symmetric spaces.
In this case, the geodesic distance corresponds to an on-shell relation invariant under the action of the (deformed) relativistic symmetries, and the orbit corresponding to a given value of the mass $m$ is spanned by the action of (deformed) Lorentz transformations.
Explicitly, the on-shell relation for a particle of mass $m$ and momentum $p_{\mu}$ is given by
\begin{equation}
m=D(0,p_{\mu})=\int_{0}^{1}ds\sqrt{g^{\mu\nu}\left(\gamma\right)\dot{\gamma}_{\mu}\dot{\gamma}_{\nu}}, \qquad \gamma(0)_{\mu}=0, ~~ \gamma(1)_{\mu}=p_{\mu}
\label{disp-rel-diff}
\end{equation}
 where $\gamma_{\mu}(s)$ is the solution of the geodesic equation
\begin{equation}
\ddot{\gamma}_{\mu}+\Delta_{\mu}^{\ \nu\rho}\dot{\gamma}_{\nu}\dot{\gamma}_{\rho}=0\ ,\label{geodesicEq}
\end{equation}
 $\Delta_{\mu}^{\ \nu\rho}$ being the Christoffel symbols of the metric $g^{\mu\nu}$, defined by the
relation\footnote{Notice the role of the upper and lower indexes, due to the momentum-space metric, and such that $\partial^\mu = \partial/\partial p_\mu$, where we used the common notation for partial derivatives $f^{,\mu} = \partial^\mu f$.}
\begin{equation}
\Delta_{\mu}^{\ \nu\rho}=\frac{1}{2}g_{\mu\sigma}\left(g^{\sigma\nu,\rho}+g^{\sigma\rho,\nu}-g^{\nu\rho,\sigma}\right) ,
\label{christoffel}
\end{equation}
$s$ is an affine parameter on the curve $\gamma(s)$, and the dot stands for the derivative with respect to $s$: $\dot{\gamma}_\mu = d\gamma_\mu/ds$.

Besides the deformed on-shell relation, the curvature of momentum space manifests itself in a deformed  composition law $\left( p \oplus q \right)$ between different particles momenta.
Following Ref.~\cite{RelLocPrinciple} the composition law determines a connection on $\cal{P}$ that, in the origin of momentum space, takes the form
\begin{equation}
\Gamma_\mu^{\rho \sigma} (0) = - \frac{\partial}{\partial p_\rho} \frac{\partial}{\partial q_\sigma} \left( p \oplus q \right)_\mu\Big|_{p,q=0}.
\label{connectionOrigin}
\end{equation}

\begin{figure}[htbp]
\includegraphics[scale=0.6]{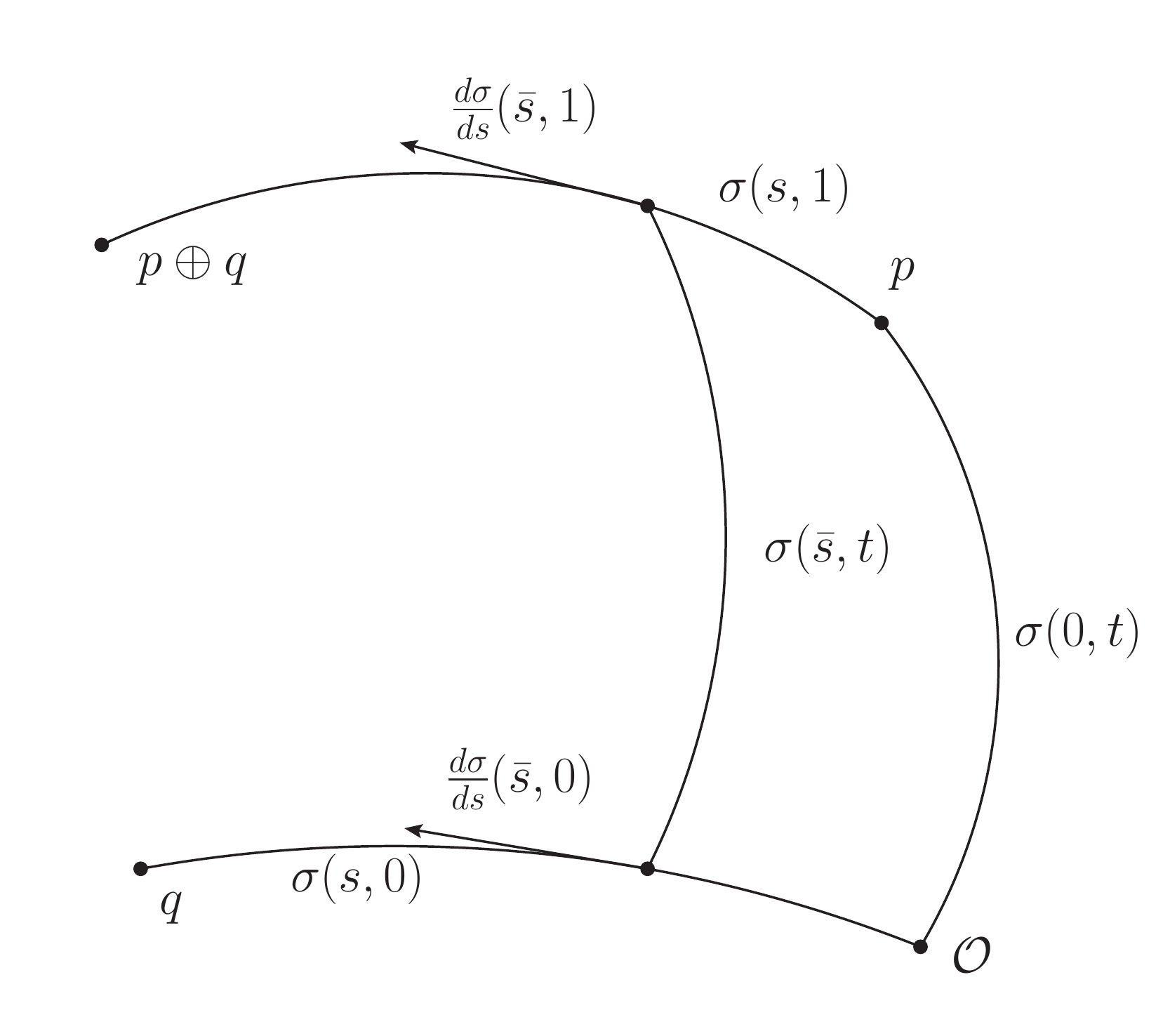}
\caption{In this figure we show a visualization of the construction of the composition law from the connection geodesics proposed in~\cite{palmisano}.
To each of the two momentum space points $p\equiv p({\cal P})$ and $q\equiv q({\cal Q})$ one associates respectively the connection geodesics $\sigma^{(p)}$
and $\sigma^{(q)}$ that connect them to the origin of momentum space.
One then defines a parametric surface $\sigma(s,t)$ with boundary conditions determined by the curves $\sigma^{(p)}(s)$ and $\sigma^{(p)}(t)$ and defined by imposing the tangent vector $\frac{d\sigma}{ds}(s,t)$ to be parallel transported along the integral curves associated to the tangent vector $\frac{d\sigma}{dt}(s,t)$. The composition law is then defined as the extremal point $p\oplus q = \sigma(1,1)$.}
\label{fig:composition}
\end{figure}

An alternative description of the relationship between composition law and affine connection
was proposed in Ref.~\cite{palmisano} (also see the preliminary proposal by Mercati~\cite{flaviotalk}).
 The geometrical interpretation proposed in Ref.~\cite{palmisano} is illustrated in our Fig.~\ref{fig:composition}: each point $P$ is associated to the connection geodesic $\sigma^{(P)}(s):[0,1] \rightarrow \cal{P}$ (relative to the connection $\Gamma$) which connects it to the origin of momentum space, i.e. $p_\mu(P)=\sigma^{(P)}_\mu(1)$.
Considering two points $P,\ Q$ with coordinates $p\equiv p(P),\ q\equiv q(Q)$, one defines a parametric surface $\sigma(s,t):[0,1]\times[0,1]\rightarrow\cal{P}$ with boundary conditions $\sigma(s,0) = \sigma^{(P)}(s)$, $\sigma(0,t) = \sigma^{(Q)}(t)$, and defined so that at any point of the surface the tangent vector $\frac{d\sigma}{ds}(s,t)$ is parallel transported along the integral curves associated to the tangent vector $\frac{d\sigma}{dt}(s,t)$, i.e. satisfying the condition
\begin{equation}
\frac{d\sigma_\mu(s,t)}{dt}\nabla^\mu_\Gamma \frac{d\sigma_\nu(s,t)}{ds}=0,
\label{parallelTransport}
\end{equation}
with $\nabla_\Gamma^\mu V_\nu(P) = \partial^\mu V_\nu(P) + \Gamma_\nu^{\mu\rho}(P) V_\rho(P)$ the covariant derivative associated to $\Gamma$.
The composition law for two momenta is then defined as the extremal point of $\sigma(s,t)$:
\begin{equation}
p\oplus q = \sigma(1,1).
\label{ompositionGamma}
\end{equation}

The construction can be extended to define a ``translated'' composition law $\oplus^{[P]}$, where any momentum is associated to the geodesic connecting it to the ``subtraction point'' $P$ instead of the origin of momentum space.
In this case, at second order in the curvature scale, the connection at $P$ and the composition law satisfy the relation
\begin{equation}
\Gamma_\mu^{\rho \sigma} (P) = - \frac{\partial}{\partial p_\rho} \frac{\partial}{\partial q_\sigma} \left( p \oplus^{[k(P)]} q \right)_\mu\Big|_{p,q=k(P)}.
\label{connectionP}
\end{equation}
In the original formulation~\cite{RelLocPrinciple} of relative locality, a different formula for the ``translated'' composition law was adopted, namely 
\begin{equation}
p \oplus^{[k(P)]} q = k \oplus \left( \left( \ominus k \oplus p \right) \oplus \left( \ominus k \oplus q \right) \right),
\end{equation}
the associated connection still being related by expression (\ref{connectionP}).

The advantage of the geometrical interpretation proposed in~\cite{palmisano} is that it allows to associate (univocally) a composition law to a given connection, $\Gamma \rightarrow \oplus$, which is not the case for the original proposal of the ``translated'' composition law defined in~\cite{RelLocPrinciple}.
However, one can easily see (see~\cite{CarmonaRelancioComposition}) that if the composition law is associative, i.e. if $p\oplus(q\oplus k) = (p\oplus q)\oplus k $, then the two definitions coincide. This is the case for the explicit examples that we will consider, based on different realizations of $\kappa$-momentum space which, corresponding to the Lie group manifold $AN_3$ (see Sec.~\ref{sec:time-order}), is characterized by an associative composition law.

Finally we point out that the composition law must obey certain compatibility conditions with the metric in order to achieve a genuinely relativistic framework.
In particular one requires the summation law to be compatible with the generalized (deformed) Lorentz transformation (see below).
While we will assume in the following of this manuscript that these relativity conditions are fulfilled, we will make them explicit when considering an example of maximally symmetric momentum space based on the $\kappa$-Poincar\'e/$\kappa$-Minkowski description of symmetries.

Before passing to the investigation of the behaviour under diffeomorphisms of the specific components of the relative locality framework, let us first review briefly how these structures are implemented in the relative locality action.
When describing the kinematics of a system of interacting particles, for each particle the action will have a bulk term
\begin{equation}
\int \left( {\cal L}_{kin}(p) - N \left(D(0, p)^2-m_p^2\right)\ \right),
\end{equation}
where $N$ is a Lagrange multiplier imposing the on-shell relation, and the kinetic term is
\begin{equation}
{\cal L}_{kin}(p)=\chi^\mu_p \dot{p}_\mu,
\end{equation}
for some spacetime coordinates $\chi^\mu$ tangent to momentum space.
For each interaction the action will have a boundary term
\begin{equation}
\zeta^\mu {\cal K}_\mu(p,q,\dots) \ ,
\end{equation}
where $\zeta^\mu$ is a Lagrange multiplier imposing the conservation law ${\cal K}_\mu(p,q,\dots) =0$, $p,q,\dots$ being the momenta of the particle involved in the chain of processes.
${\cal K}_\mu(p,q,\dots) $ here is a function of the momenta constructed from the composition law $\oplus_\Gamma$, where we emphasize the fact that $\oplus$ is associated with a certain connection $\Gamma$.
The total action will look like
\begin{equation}
\sum_i \int \left( {\cal L}_{kin}(p_i) - N \left(D(0, p_i)^2-m_{p_i}^2\right)\ \right) + \sum_I \zeta^\mu_I{\cal K}^I_\mu(p_1,p_2,\dots),
\end{equation}
where $i$ sums over particles, and $I$ over the number of vertices.

We postpone a more detailed discussion of the boundary terms to Sec.~(\ref{sec:boundary}). First, we will focus on the role of the on-shell relation and its behaviour under (passive) diffeomorphisms.

\subsection{On-shell relation and momentum-space diffeomorphisms}
\label{onshell-diff-sec}
Let us consider two sets of momentum-space coordinates
$p_{\mu}$ and $\tilde{p}_{\nu}$  related by the passive diffeomorphism (change of coordinates)
\begin{equation}
p_{\mu}=f_{\mu}(\tilde{p}_{\nu})\ .\label{change}
\end{equation}
Substituting the map (\ref{change})
in the on-shell relation (\ref{disp-rel-diff}), written in coordinates $p_{\mu}$,
one directly finds (see also~\cite{palmisano})
\begin{equation}
m=D(0, p_{\mu})=D(0, f_{\mu}(\tilde{p}_{\nu}))=D(0, \tilde{p}_{\mu}).
\label{on-shell-diff}
\end{equation}
By inserting the passive diffeomorphism in the on-shell relation, one actually gets the
on-shell relation in the new coordinates $\tilde{p}_{\nu}$. To see this, it is sufficient to note that the integrand in the geodesic distance in (\ref{disp-rel-diff}) is a
scalar function of $p_{\mu}$. Indeed the $\dot{\gamma}_\mu$ transform as a vector
\begin{equation}
\dot{\gamma}_\mu = \frac{d f_\mu (\tilde{\gamma})}{ds} = \frac{\partial f_\mu (\tilde{\gamma})}{\partial \tilde{\gamma}_\nu} \dot{\tilde{\gamma}}_\nu,
\label{scalar1}
\end{equation}
while the metric tensor transforms as a two-rank tensor
\begin{equation}
g^{\mu\nu}(\gamma) = \frac{\partial f^{-1}_\rho (\gamma)}{\partial \gamma_\mu} \frac{\partial f^{-1}_\sigma (\gamma)}{\partial \gamma_\nu} \tilde{g}^{\rho\sigma}(\tilde{\gamma}) ,
\label{scalar2}
\end{equation}
where $f^{-1}_\mu(\gamma) = \tilde{\gamma}_\mu$. By the chain rule
\begin{equation}
\frac{\partial f^{-1}_\mu (\gamma)}{\partial \gamma_\rho} \frac{\partial f_\rho (\tilde{\gamma})}{\partial \tilde{\gamma}_\nu} = \frac{\partial f^{-1}_\mu (\gamma)}{\partial \tilde{\gamma}_\nu} = \delta^\nu_\mu,
\label{scalar3}
\end{equation}
so that from (\ref{scalar1}),(\ref{scalar2}) and (\ref{scalar3}) one gets
\begin{equation}
g^{\mu\nu} \dot{\gamma}_\mu \dot{\gamma}_\nu = \tilde{g}^{\mu\nu} \dot{\tilde{\gamma}}_\mu \dot{\tilde{\gamma}}_\nu,
\label{scalar4}
\end{equation}
which implies Eq.~(\ref{on-shell-diff}), and which can be also ascribed to the fact that the geodesic distance from the origin depends only on the final point.

Notice that this means that the on-shell relation in coordinates $\tilde{p}_{\nu}$, can be obtained directly starting from the on-shell relation in coordinates $p_{\mu}$, and rewriting it in terms of the coordinates $\tilde{p}_{\nu}$ through the map (\ref{change}).
We will return to this obvious but important point in Sec.~\ref{sec:on-shell-k-bases}, showing its behaviour for an explicit example of momentum space.

\subsection{Including boundary terms}
\label{sec:boundary}

Besides the onshell relation, a relative-locality model is specified by the boundary terms that govern the interaction processes~\cite{RelLocPrinciple}. A boundary term enforces on one hand the momenta conservation law at the vertex, and on the other hand it generates translations through its action by Poisson brackets (see~\cite{RelLocPrinciple,anatomy}, and App.~\ref{app:translationsGeneric}).
It is worth noting that starting from a definite set of coordinates in momentum space, the same conservation law for a certain process can be implemented by several different choices of the boundary term. Indeed, in general, given the conservation law
\begin{equation}
{\cal K}_\mu(p,q,\dots)= 0,
\label{ConservationImplicit}
\end{equation}
one can define a wide class of alternative boundary terms ${\cal K}'_\mu$ which satisfy the same conservation law~(\ref{ConservationImplicit}).
On the other hand, given a set of coordinates in momentum space, which in turn fixes the Poisson brackets $\{p_\mu,x^\nu\}$ between momenta and spacetime coordinates, the translations generated by ${\cal K}_\mu$ and ${\cal K}'_\mu$ are in general different. In this section the discussion is kept as general as possible, but the physical content will become clearer in the following sections when we will focus on an explicit example of momentum space.

The conservation law is related to a deformed summation rule $\oplus$, which must be compatible to the deformed symmetries of the theory.
Taking for instance a process with two incoming ($p,q$) and one outgoing $k$ particle, one has the conservation law
\begin{equation}
\left( p \oplus q  \right)_\mu = k_\mu,
\label{conservation}
\end{equation}
where $\oplus$ encodes the deformation in the summation rule of momenta, which in general is itself a function of the momenta involved in the process.

The conservation law $\ref{conservation}$ is justified by the fact that the quantity $( p \oplus q )$ transforms, under generic diffeomorphisms, with the same law of a momentum-space coordinate, thus providing a good definition for the ``total momentum'' of the system of particles with momenta $p$ and $q$.
The change of $( p \oplus q )$ under diffeomorphisms can be deduced relying on the construction outlined in Sec.~\ref{sec:RelLoc}. In particular (see~\cite{palmisano}) one can show that, given the (passive) diffeomorphism (\ref{change}) ($p_\mu = f_\mu(\tilde{p})$), the following property holds
\begin{equation}
(p\oplus q)_\mu \equiv (p\oplus_\Gamma q)_\mu = (f(\tilde{p})\oplus_{f(\tilde{\Gamma})} f(\tilde{q}))_\mu=f_\mu(\tilde{p} \oplus_{\tilde{\Gamma}} \tilde{q}) \equiv f_\mu(\tilde{p}\tilde{\oplus} \tilde{q}),
\label{changeOplus}
\end{equation}
where, with self-explanatory notation, we have taken into account of the change of the composition law $\oplus \rightarrow \tilde{\oplus}$ due to the transformation law $\Gamma = f(\tilde{\Gamma})$ of the connection under diffeomorphisms
\begin{equation}
\Gamma_\mu^{\rho\sigma}(k) = f(\tilde{\Gamma}(\tilde{k}))_\mu^{\rho\sigma}
=\bar{\cal{M}}_\alpha^\rho\bar{\cal{M}}_\beta^\sigma{\cal{M}}_\mu^\lambda \tilde{\Gamma}_\lambda^{\alpha\beta}(\tilde{k})- \bar{\cal{M}}_\alpha^\rho\bar{\cal{M}}_\beta^\sigma\partial^\beta{\cal{M}}_\mu^\alpha ,
\label{transformGamma}
\end{equation}
with ${\cal{M}}_\mu^\alpha(\tilde{k})= \partial f_\mu(\tilde{k})/\partial \tilde{k}_\alpha$, and $\bar{\cal{M}}^\mu_\alpha(\tilde{k})$ its inverse.
Indeed, using (\ref{transformGamma}) and the fact that $d\sigma_\mu/ds$ transforms as a vector, $d\sigma_\mu/ds = {\cal{M}}_\mu^\alpha d\tilde{\sigma}_\alpha/ds$, where the surface is mapped as $\sigma(s,t) = f(\tilde{\sigma}(s,t))$, one can verify that Eq.~(\ref{parallelTransport}) implies
\begin{equation}
\frac{d\tilde{\sigma}_\mu(s,t)}{dt}\nabla^\mu_{\tilde{\Gamma}} \frac{d\tilde{\sigma}_\nu(s,t)}{ds}=0.
\end{equation}
Thus, the surface $\tilde{\sigma}$ is a solution of the parallel transport equation with connection $\tilde{\Gamma}$ and boundary points $\tilde{p} = \tilde{\sigma}(1,0)$ and $\tilde{q} = \tilde{\sigma}(0,1)$, and we get, in particular, that $p\oplus q = \sigma(1,1) = f(\tilde{\sigma}(1,1))=f(\tilde{p}\oplus \tilde{q})$.

Eq.~(\ref{changeOplus}) expresses that $(p\oplus q)_\mu$ transforms under diffeomorphisms with the same law of a single momentum $p_\mu$ (thus $(p\oplus q)$ can be interpreted as a total momentum).
This guarantees in particular that, in the case of a maximally symmetric momentum space, under (deformed) Lorentz transformations, which are a subset of the possible diffeomorphisms, the conservation law (\ref{conservation}) transforms covariantly. Indeed, denoting as $\Lambda (p)$ the deformed Lorentz transformation\footnote{For a maximally symmetric momentum space the deformed Lorentz transformations correspond to the charge/generators associated to the Killing vectors of the metric that reduce to standard Lorentz transformations in the limit of vanishing momentum space curvature.} associated to a specific maximally symmetric momentum space, Eq.~(\ref{changeOplus}) implies that, if an observer describes the summation law $p\oplus q$, a second observer, relatively boosted (or rotated) respect to the first, describes the same quantity as
\begin{equation}
\left(\Lambda(p)\oplus_{\Lambda(\Gamma)} \Lambda(q)\right)_\mu = \Lambda_\mu(p\oplus_\Gamma q) .
\label{LorentzOplus}
\end{equation} 
In turn, last expression ensures that the conservation law for a given process is preserved under boosts, so that if it holds for the first observer, it holds also for the relatively boosted (or rotated) one.
Taking again as example the conservation law~(\ref{conservation}), the l.h.s. of~(\ref{conservation}) transforms as $\Lambda_\mu(k)$, while the r.h.s. as in~(\ref{LorentzOplus}), i.e.~(\ref{conservation}) transforms as
\begin{equation}
k_\mu  = \left( p \oplus_\Gamma q  \right)_\mu \quad \Rightarrow \quad \Lambda_\mu(k) = \Lambda_\mu(p\oplus_\Gamma q) = \left(\Lambda(p)\oplus_{\Lambda(\Gamma)} \Lambda(q)\right)_\mu.
\end{equation}

Thus, if the summation law is built from a connection (obeying transformation law~(\ref{transformGamma})), and the momentum space is maximally symmetric, the invariance under (deformed) Lorentz transformations is ensured.
The converse is also obviously true, considering the notion of parallel transport at the basis of the construction outlined in Sec.~\ref{sec:RelLoc}: if one can define a summation law for momenta such that the composition of two momenta transforms with the same law of a single momentum coordinate, then this automatically defines a connection in momentum space through relation (\ref{connectionP}).
The requirement of covariance however does not single out the summation law or the associated connection, unless some further restriction is imposed, like for instance the requirement for the sum to be associative, or eventually some restriction on the torsion and curvature of the connection (see the discussions in~\cite{palmisano}, in~\cite{GACboosts} and in~\cite{CarmonaGoldenRule}). 
We will see in section~\ref{sec:time-order} for the example of $\kappa$-momentum space, that relying on the structures given by the Hopf-algebras, a covariant summation law arises naturally, which, due to the underlying group structure, is associative (but not commutative), and defines a non-metric, torsionful, connection. Another possibility, which was explored in~\cite{palmisano}, would be to start from a metric connection. As it was shown in~\cite{palmisano}, the associated summation law is not associative in that case.

Once specified the summation law relative to a set of momentum space coordinates, the conservation law (\ref{conservation}) can be implemented by different choices of the boundary term.
We will adopt in this manuscript the prescription for which one chooses, between the class of admissible boundary terms enforcing the conservation law~(\ref{conservation}), the one corresponding to
\begin{equation}
{\cal K}_\mu=P_{\mu}^{\text{ in}}-P_{\mu}^{\text{ out}}
\label{totalMomInOut}
\end{equation}
where $P_{\mu}^{\text{ in}}$ and $P_{\mu}^{\text{ out}}$ are respectively the total momentum incoming and outgoing the interaction, computed with the deformed composition law $\oplus$. As it has been shown in~\cite{anatomy} this choice of boundary terms is compatible with the definition of translational symmetry in the theory, and its compatibility with Lorentz (boost) transformations has been recently established in~\cite{GiuliaBoosts}.
Specifically the notion of translational symmetry implemented in~\cite{anatomy} is consistent with the requirement that all the lagrange multipliers $\zeta^\mu$, which turn out to play the role of ``interaction coordinates'', change under translation by the same amount $b^\mu$, i.e. $\delta z^\mu = b^\mu$ (see also App.~\ref{app:translations}).

For the particular case (\ref{conservation}) of two incoming and one outgoing particles, the prescription (\ref{totalMomInOut}) amounts to the expression\footnote{We are
not considering here the alternative choice of boundary terms of the form ${\cal K}_\mu = \left(\left( p \oplus q \right) \oplus \left( \ominus k \right)\right)_\mu  $, which was first proposed in~\cite{RelLocPrinciple}, and then used in some studies~\cite{FreSmoGRBrelLoc,FreidelSnyderRelLoc} due to its geometrical properties, since it was shown in~\cite{anatomy} that it is not compatible with a well-defined notion of translational symmetry.}
\begin{equation}
{\cal K}_\mu = \left( p \oplus q \right)_\mu - k_\mu.
\label{boundaryProper}
\end{equation}
To some extent this choice looks the most ``natural'', as it does not imply extra functions of momenta multiplying the incoming or outgoing total momentum, and we will denote the theory built on this prescription~\footnote{Alternatives to our prescriptions may be worthy of investigation. We postpone their discussion to the final section~\ref{sec:conclusions}.} as the ``proper'' theory $S(p)$, indicating with $S(p)$ the action constructed with the prescription~(\ref{boundaryProper}), starting from the coordinates $p$ on momentum space ${\cal{P}}$.
However, any boundary term of the form
\begin{equation}
{\cal K}^{\cal F}_\mu = {\cal F}_\mu \left( {\cal K} \right)= {\cal F}_\mu \left( p \oplus q \right) - {\cal F}_\mu (k),
\label{boundaryAlternative}
\end{equation}
where ${\cal F}_\mu$ is an invertible map , will produce the conservation law (\ref{conservation}).
On the other hand, the translations generated by ${\cal K}_\mu$ and ${\cal K}^{\cal F}_\mu$ are in general different:
\begin{equation}
\left\lbrace {\cal K}_\mu , \cdot \right\rbrace =
\left\lbrace (p \oplus q )_\mu , \cdot \right\rbrace - \left\lbrace k_\mu , \cdot \right\rbrace,
\label{translationProper}
\end{equation}
\begin{equation}
\left\lbrace {\cal K^{\cal F}}_\mu , \cdot \right\rbrace = \frac{\partial{\cal F}_\mu (p \oplus q )}{\partial (p \oplus q )_\rho} \left\lbrace (p \oplus q )_\rho , \cdot \right\rbrace - \frac{\partial{\cal F}_\mu (k)}{\partial k_\rho} \left\lbrace k_\rho, \cdot \right\rbrace  \neq \left\lbrace {\cal K}_\mu , \cdot \right\rbrace
\label{translationAlternative}
\end{equation}

Notice at this point that defining the boundary term as in (\ref{boundaryProper}), it follows from (\ref{changeOplus}) that under diffeomorphisms the boundary term changes as
\begin{equation}
 {\cal K}_\mu = (p\oplus q)_\mu-k_\mu = (f(\tilde{p})\oplus f(\tilde{q}))_\mu-f_\mu(\tilde{k}) = f_\mu(\tilde{p}\tilde{\oplus} \tilde{q}) - f_\mu (\tilde{k}) = \tilde{{\cal K}}^f_\mu \neq \tilde{{\cal K}}_\mu ,
 \label{boundaryDiffeo}
\end{equation}
where $\tilde{{\cal K}}_\mu=\tilde{p}\tilde{\oplus}\tilde{q}_\mu- \tilde{k}_\mu$ is the boundary term one would have chosen to build the proper theory in coordinates $\tilde{p}_\mu$, i.e. the boundary term obtained following the prescription (\ref{boundaryProper}), starting from a theory defined in momentum space coordinates $\tilde{p}_\mu$.
Notice also that the relation between the boundary terms $\tilde{{\cal K}}_\mu$ and $\tilde{{\cal K}}^f_\mu$ is of the same kind of the one described by Eqs.~(\ref{boundaryProper}) and~(\ref{boundaryAlternative}), so that, as shown in Eqs.~(\ref{translationProper})-(\ref{translationAlternative}), in general they generate the same conservation laws but different translation transformations.
Thus, the diffeomorphism (\ref{change}) does not map the theory $S(p_\mu)$, the proper theory in coordinates $p_\mu$ obtained with the ``natural'' prescription (\ref{boundaryProper}) for the boundary terms, to the proper theory $S(\tilde{p}_\mu)$ in coordinates $\tilde{p}_\mu$, the one with boundary terms $\tilde{{\cal{K}}}(\tilde{p})_\mu$, but it maps it to an ``improper'' theory $S^f(\tilde{p}_\mu)$ characterized by boundary terms of the kind $\tilde{\cal{K}}^f(\tilde{p})$. We will show in the next sections that while the theories $S(p_\mu, {\cal K}_\mu)$ and $S(\tilde{p}_\mu,\tilde{{\cal K}}_\mu)$ lead to different predictions for the observables relative to a certain process, the theories $S(p_\mu, {\cal K}_\mu)$ and $S^f(\tilde{p}_\mu,\tilde{{\cal K}}^f)$ lead to the same predictions, as it is expected since they are connected just by a reparametrization.

Denoting as ``proper'' the theories constructed with the notion of ``naturality'' discussed above, and ``improper'' the theories with boundary terms of the kind $f_\mu ( \tilde{\cal{K}})$, we can delineate the following diagram:

\begin{figure}[h!]
\begin{center}
\includegraphics[scale=1]{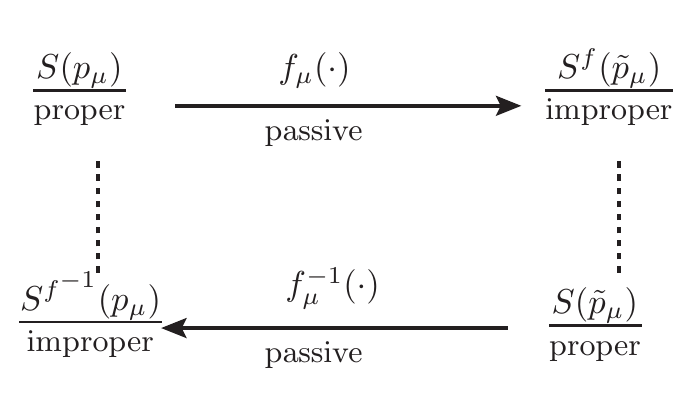}
\end{center}
\caption{The diagram shows how passive diffeomorphisms connect physically equivalent proper theories with improper theories.
On the contrary, proper theories in different coordinates, which turn out to be physically inequivalent, are not connected by passive diffeomorphisms.}
\label{fig:diagramPropImprop}
\end{figure}

\section{Diffeomorphisms between ``time-ordered'' $\kappa$-momentum spaces}

In order to clarify the considerations of the previous section, we will now focus on a specific example of relativistic curved momentum space, characterized by a de Sitter type of curvature. As stated before the maximal symmetry of de Sitter space ensures the existence of the whole set of relativistic symmetries, and in particular it has been shown~\cite{JurekDeSitt,JurekFrekfield1e2} that a de Sitter curved momentum space corresponds to a description of a Hopf-algebraic non-commutative spacetime of $\kappa$-Minkowski/$\kappa$-Poincar\'e type~\cite{Lukierski,MajidRuegg}, where the deformation of relativistic symmetries is encoded in the Hopf-algebra structures of $\kappa$-Poincar\'e.
We will denote the set of momentum space geometrical structures associated to this kind of spacetimes as $\kappa$-momentum space.
We will consider two choices of bases for the so-called ``time-ordered bases'' of $\kappa$-momentum space~\cite{MajidOecklkFuorier,LukKosMaskfield,gacMajid,GACagostinikfield2004}, study the relative locality action associated to each choice of basis, and consider the effect of diffeomorphisms relating the two coordinate bases.
Before doing so we discuss the construction of $\kappa$-momentum space and its relation with $\kappa$-Poincar\'e/$\kappa$-Minkowski.

\subsection{The bicrossproduct or ``time-ordered'' $\kappa$-Poincar\'e basis}
\label{sec:time-order}

The bicrossproduct basis of $\kappa$-Poincaré $\cal{P}_\kappa$ was introduced in~\cite{MajidRuegg} as the Hopf algebra deformation of special relativistic symmetries with structure $U(so(1,3))\cobicross T$, with $T$ the translation sector.
It has been noticed however~\cite{GACagostinikfield2004,MeljakBasis,BoroPachokBasis,meljanac} that the Hopf-algebra defined in~\cite{MajidRuegg}, sometimes called ``Majid-Ruegg'' basis in the literature, corresponds only to a specific choice of basis within the class of possible bicrossproduct formulations of $\kappa$-Poincaré.
We consider in this paper a generalized class of $\kappa$-Poincaré bicrossproduct bases which takes the form
\begin{equation}
\begin{gathered}
\left[P_{\mu},P_{\nu}\right]=0\ ,\qquad\left[R_{j},R_{k}\right]=\epsilon_{jkl}R_{l}\ ,\qquad\left[\mathcal{N}_{j},\mathcal{N}_{k}\right]=-\epsilon_{jkl}R_{l}\ , \\
\left[R_{j},P_{0}\right]=0\ ,\qquad\left[R_{j},P_{k}\right]=\epsilon_{jkl}P_{l}\ ,\qquad\left[R_{j},{\cal N}_{k}\right]=\epsilon_{jkl}{\cal N}_{l}\ , \\
\left[\mathcal{N}_{j},P_{0}\right]=e^{\lambda\ell P_{0}}P_{j}\ ,\qquad\left[\mathcal{N}_{j},P_{k}\right]=\delta_{jk}\left(\frac{e^{(2-\lambda)\ell P_{0}}-e^{-\lambda\ell P_{0}}}{2\ell}-\frac{\ell}{2}e^{\lambda\ell P_{0}}\vec{P}^{2}\right)+(1-\lambda)\ell e^{\lambda\ell P_{0}}P_{j}P_{k}\ ,
\end{gathered}
\label{k-Poinc}
\end{equation}
\begin{equation}
\begin{gathered}
\Delta P_{0}=P_{0}\otimes\mathbf{1}+\mathbf{1}\otimes P_{0}, \qquad \Delta P_{j}=P_{j}\otimes e^{-\lambda\ell P_{0}}+e^{(1-\lambda)\ell P_{0}}\otimes P_{j}, \\
\Delta R_{j}=R_{j}\otimes\mathbf{1}+\mathbf{1}\otimes R_{j}, \qquad
\Delta{\cal N}_{j}={\cal N}_{j}\otimes\mathbbm{1}+e^{\ell P_{0}}\otimes{\cal N}_{j}-\ell\epsilon_{jkl}e^{\lambda\ell P_{0}}P_{k}\otimes R_{l}\ ,
\end{gathered}
\label{coproducts}
\end{equation}
where $E$, $P_j$, $N_j$ and $M_j$ are respectively time translation, space translation, boost and rotation generators. Here $\ell \propto 1/M_p$ ($\ell=-1/\kappa$, in the conventions used in~\cite{MajidRuegg}) is a deformation parameter with dimensions of an inverse-momentum. 
The different bases are parametrized by $\lambda \in [0,1]$, so that for $\lambda=0$ the Majid-Ruegg basis is recovered.
The algebra (\ref{k-Poinc}) admits a quadratic invariant element (the mass Casimir)
\begin{equation}
\Box_\lambda = \left(\frac{2}{\ell}\right)^{2}\sinh^{2}\left(\frac{\ell}{2}P_{0}\right)-\vec{P}^{2}e^{\ell\left(2\lambda-1\right)P_{0}} .
\label{Casimir}
\end{equation}
While we postpone a detailed discussion to a future work, we here generalize to this class of bases some of the properties already considered for the Majid-Ruegg basis.

The bicrossproduct construction has, among others, two important properties: the Lorentz sector of the algebra is undeformed; the momenta (i.e. the translation generators) are in full duality with, and act homogeneously~\cite{MajidRuegg} on, the $\kappa$-Minkowski non-commutative coordinates, defined by
\begin{equation}
[X^0,X^j]=\ell X^j , \qquad [X^j,X^k]=0 .
\label{kMinkowski}
\end{equation}
If one consider the set of ``time-ordered'' plane waves
\begin{equation}
u_p(X)= \ :\!e^{p_\mu X^\mu}\!:\  = e^{\lambda p_0 X^0}e^{p_j X^j}e^{(1-\lambda)p_0 X^0} ,
\label{planeWaves}
\end{equation}
it's easy to see~\cite{JurekDeSitt,JurekFrekfield1e2} that these are elements of the group $AN(3)$, the Lie group corresponding to half of de Sitter space $SO(4,1)/SO(3,1)$ arising in the Iwasawa decomposition of SO(4,1).
The plane waves~(\ref{planeWaves}) form a basis for the action of ${\cal P}_\kappa$ (cf.~\cite{JurekDeSitt,gacMajid,GACagostinikfield2004}) and the momentum space, parametrized by $p_\mu$ coordinates, inherits the de Sitter geometry from the group structure of $AN(3)$ (see also~\cite{FlaGiukRelLoc}).
It is possible to show\footnote{One way of doing that is to consider, following the construction of~\cite{JurekDeSitt}, the matrix representation of the Borel group $AN(3)=SO(4,1)/SO(3,1)$ and to obtain the 5D coordinates relative to the choice of ordering~(\ref{planeWaves}). These coordinates induce on the hyperboloid the metric~(\ref{metricInterval}) (see also~\cite{MeljakBasis,meljanac}).} that for this class of ``time-ordered'' plane waves the metric in momentum space is (see also~\cite{MeljakBasis,meljanac})
\begin{equation}
dp^{2}=g^{\mu\nu}\left(p\right)dp_{\mu}dp_{\nu}=\left(1-\lambda^{2}\ell^{2}e^{-2\ell\left(1-\lambda\right)p_{0}}\vec{p}^{2}\right)dp_{0}^{2}-e^{-2\ell\left(1-\lambda\right)p_{0}}d\vec{p}^{2}-2\lambda\ell e^{-2\ell\left(1-\lambda\right)p_{0}}\vec{p}\cdot d\vec{p}dp_{0}.
\label{metricInterval}
\end{equation}
One can show that this metric is invariant under the action of deformed Lorentz transformations of momenta (as shown for the Majid-Ruegg case in~\cite{FlaGiukRelLoc}).

The full duality between momenta and coordinates of the bicrossproduct construction can be appreciated by noticing that the translation coproducts~(\ref{coproducts}) can be derived from the group product law, defining the action~\cite{GACagostinikfield2004}
\begin{equation}
P_\mu \triangleright u_p(X) = p_\mu u_p(X), \qquad P_\mu \triangleright u_p(X) \cdot u_q(X) = \cdot (\Delta P_\mu (u_p(X) \otimes u_q(X)) )
\end{equation}
on the product of plane waves
\begin{equation}
u_p(X) \cdot u_q(X) = u_{p\oplus q}(X) .
\label{oplus}
\end{equation}
From the definitions~(\ref{oplus}) and~(\ref{planeWaves}), using Eq.~(\ref{kMinkowski}) one gets (for instance using the Baker-Campbell-Hausdorff formula for the product of exponentials of noncommutative Lie-algebra elements)
\begin{equation}
(p\oplus q)_0 = p_0 + q_0, \qquad
(p\oplus q)_j = p_j e^{-\lambda q_0} + e^{(1-\lambda)\ell p_0} q_0 ,
\label{sumTorder}
\end{equation}
which show, comparing with~(\ref{coproducts}), the correspondence of the description of momenta as functions $p_\mu$ on the group $AN(3)$ and as generators of translations $P_\mu$ in ${\cal P}_\kappa$.
Following the construction outlined in Sec.~\ref{sec:RelLoc}, one can show that the composition law (\ref{sumTorder}) can be associated to a connection
\begin{equation}
 \Gamma_\mu^{\rho\sigma} (k) = \delta_\mu^j\left( \delta^\rho_0 \left( \lambda-1 \right) \left( \delta^\sigma_j \ell + \delta^\sigma_0 \lambda \ell^2 k_j \right) + \delta^\rho_j \delta^\sigma_0 \lambda \ell \right),
 \label{connectionk}
\end{equation}
which can be shown to be non-metric and torsionful.
As mentioned in Sec.~\ref{sec:boundary} Eq.~(\ref{oplus}) defines a deformed summation law of momenta which, thanks to the symmetry properties of the construction, in kinematical terms encodes a (deformed) Lorentz invariant energy-momentum conservation law~\cite{DSR,GACboosts}.
Indeed, the action of the boost generator, defined by the action encoded in the commutators~(\ref{k-Poinc}), on the plane wave $u_p(X)$, extends automatically to the product plane wave~(\ref{oplus}) through the coproduct action~(\ref{coproducts}), so that the composed momentum $p\oplus q$ transforms, for a finite Lorentz transformation~\cite{FlaGiukRelLoc,GiuliaBoosts}, with the same law of a single momentum:
\begin{equation}
(p\oplus q)_\mu \rightarrow (p\oplus q)'_\mu = \Lambda^\xi_\mu (p\oplus q).
\end{equation} 
Notice that in this case the coproduct structure is such that the action of the boost generator can be formulated in terms of a total boost~\cite{GACboosts} that obeys the decomposition
\begin{equation}
\Lambda^\xi_\mu (p\oplus q) =  (\Lambda^\xi(p)\oplus \Lambda^{\xi\triangleleft p}(q))_\mu,
\end{equation} 
where $\xi\triangleleft p$ denotes the well-known ``backreaction'' from the momentum of the first particle (see~\cite{FlaGiukRelLoc,GiuliaBoosts}). 
From the considerations of Sec.~\ref{sec:boundary}, this can be in turn attributed to the non-metricity~\footnote{If one had started, as explored in~\cite{palmisano}, from a metric (Levi-Civita) connection ${\cal A}_\mu^{\rho\sigma}$, which is invariant ($\Lambda^\xi({\cal A}) = {\cal A}$), one would have had the transformation $\Lambda^\xi_\mu (p\oplus q) =  (\Lambda^\xi(p)\oplus \Lambda^{\xi}(q))_\mu$. However the composition law would have been non-associative~\cite{palmisano}.} of the connection~(\ref{connectionk}), i.e., making explicit the role of $\Gamma$,
\begin{equation}
(\Lambda^\xi(p)\oplus_\Gamma \Lambda^{\xi\triangleleft p}(q))_\mu = (\Lambda^\xi(p)\oplus_{\Lambda^\xi(\Gamma)} \Lambda^{\xi}(q))_\mu.
\end{equation}

In the bicrossproduct construction here outlined the $\kappa$-Minkowski non-commutative coordinates generate ``translations'' in the momentum space manifold\footnote{These are not to be confused with physical translation in spacetime, but again are expression of the duality between de Sitter momentum space and $\kappa$-Poincaré, i.e. they have the same mathematical structure of translations in de Sitter spacetime if one changed momenta with spacetime coordinates.}, i.e. they generate the translational symmetries of de Sitter~\cite{JurekFrekfield1e2}.
In order to view this we notice that one can derive the killing vectors $\xi^\mu_\nu \equiv \xi_\nu({X^\mu})$ associated with ``translations'' in momentum space generated by $X^\mu$, satisfying the Killing equation\footnote{The covariant derivative is defined from the metric (\ref{metric}) as $\nabla^\mu V^\nu = \partial^\mu V^\nu - \Delta_\rho^{\ \mu\nu} V^\rho $, where the Christoffel symbols $\Delta_\rho^{\ \mu\nu}$ are defined in (\ref{christoffel}).}
\begin{equation}
\nabla^\mu \xi^\nu (X)+ \nabla^\nu \xi^\mu (X)= 0.
\end{equation}
One finds
\begin{equation}
\xi^0_\mu = (1,(1-\lambda)\ell p_j) , \qquad \xi^j_\mu = (0,e^{-\lambda \ell p_0}\delta_{jk}) .
\label{killingTranslation}
\end{equation}
From this construction arises a natural definition of phase space, with spacetime coordinates defined as the functions
\begin{equation}
 x^\mu = \xi^\mu_\nu (p) \chi^\nu,
 \label{killingCoordinates}
\end{equation}
generating translations in momentum space, where $\chi^\mu$ are vectors tangent to $p_\mu$ satisfying the canonical Poisson brackets
\begin{equation}
\left\lbrace \chi^\mu , \chi^\nu \right\rbrace = 0,\qquad \left\lbrace p_\mu , \chi^\nu\right\rbrace = \delta^\nu_\mu .
\label{canonicalPB}
\end{equation}
From the last relations and~(\ref{killingCoordinates}) it follows that the Poisson Brackets are
\begin{equation}
\begin{gathered}
\left\lbrace x^0,x^j\right\rbrace = \ell x^j, \qquad \left\lbrace x^i,x^j\right\rbrace = 0, \qquad \left\lbrace p_\mu , p_\nu \right\rbrace = 0 , \\
\left\lbrace p_0, x^0 \right\rbrace = 1 , \qquad \left\lbrace p_0, x^j \right\rbrace = 0, \\
\left\{ p_{j},x^{0}\right\} = \left(1-\lambda\right)\ell p_{j}\ ,\qquad
\left\{ p_{j},x^{k}\right\} = e^{-\lambda \ell p_{0}}\delta_j^k.
\label{phaseSpace}
\end{gathered}
\end{equation}
The coordinates $x^\mu$ thus correspond to the $AN(3)$ generators $X^\mu$, and one can check that the metric~(\ref{metricInterval}) is invariant under translations generated by $x^\mu$ by Poisson brackets as
\begin{equation}
\delta p_\mu  = \epsilon_\nu \lbrace x^\nu, p_\mu \rbrace ,
\end{equation}
for some constant vector $\epsilon_\mu$.

The symplectic structure defined by~(\ref{phaseSpace}) is associated with an action
\begin{equation}
\int ds {\cal L}_{kin}(s) \qquad \text{with} \qquad {\cal L}_{kin}
= \chi^\mu \dot{p}_\mu
= x^\mu \bar{\xi}_\mu^\nu(p) \dot{p}_\nu ,
\end{equation}
where the dot stands for the derivative with respect to the parameter $s$ on the curve in momentum space, $\dot{p}_\mu = dp_\mu/ds$, and $\bar{\xi}_\mu^\nu(p)$ are the inverse of the Killing vectors (\ref{killingCoordinates})
\begin{equation}
\bar{\xi}^\mu_a = \left( \xi^{-1} \right)^\mu_a \quad \longrightarrow \quad
\bar{\xi}^0_a = (1,-(1-\lambda)\ell p_j e^{\lambda \ell p_0} ), \quad
\bar{\xi}^j_a = (0,e^{\lambda \ell p_0}\delta_{jk}).
\end{equation}
The kinetic term is thus
\begin{equation}
{\cal L}_{kin}	 =  x^0 \dot{p}_0 - x^j (1-\lambda)\ell p_j e^{\lambda \ell p_0} \dot{p}_0 + x^j e^{\lambda \ell p_0} \dot{p}_j .
\label{kineticTerm}
\end{equation}
Alternatively the same symplectic structure (\ref{phaseSpace}) and kinetic term (\ref{kineticTerm}) can be obtained through the Kirillov construction~\cite{Kirillov}
\begin{equation}
{\cal L}_{kin} = \left\langle \tilde{X} , u_p(X)^{-1} du_p(X) \right\rangle,
\end{equation}
where $\langle \cdot,\cdot \rangle$ denotes the canonical pairing between the $AN(3)$ Lie algebra and its dual linear space in the basis $\tilde{X}_\mu$
\begin{equation}
\left\langle \tilde{X}_\mu , X^\nu \right\rangle = \delta_\mu^\nu,\qquad \tilde{X}=x^\mu \tilde{X}_\mu.
\end{equation}

We have thus shown the correspondence of the class of ``time-ordered'' bicrossproduct bases of ${\cal P}_\kappa$ and the de Sitter or $\kappa$-momentum space. In the following sections we will use this construction to define the relative locality action for our class of theories.
We conclude this section by noticing that starting from the Majid-Ruegg basis ($\lambda=0$) the other time-ordered bases ($\lambda \in [0,1]$) can be obtained through a non-linear redefinition of the space translation generators
\begin{equation}
P_j \rightarrow e^{-\lambda \ell P_0} P_j
\end{equation}
to which corresponds the change of coordinates in momentum space
\begin{equation}
p_j \rightarrow e^{-\lambda \ell p_0} p_j .
\label{mapTorder}
\end{equation}
All the structures obtained in this section can be obtained by taking into account this change of coordinates.

\subsection{Aside on the on-shell relation for ``time-ordered'' bases}
\label{sec:on-shell-k-bases}

Before considering two specific ``time-ordered'' bases, we discuss the role of the on-shell relation for the generic ``time-ordered'' parametrization (\ref{metricInterval}) of $\kappa$-momentum space.
We have already shown in general that the on-shell relation, corresponding to the geodesic distance in momentum space from the origin to a given point $P$ on the mass shell orbit, changes accordingly, after a diffeomorphism, so that its expression in terms of coordinates changes in the required way for its value, the particle mass, to remain the same.
Some claims in Ref.~\cite{meljanac} provide an invitation to analyze
this result in very explicit way.
Indeed in~\cite{meljanac} the same class of momentum space metrics (\ref{metricInterval}) were considered, but different conclusions were reached.

Referring for the details of the derivation to App.~\ref{sec:on-shell}, we here mention some steps useful to clarify the following discussion.
To find the on-shell relations relative to the class of metrics~(\ref{metricInterval}) we have to evaluate the geodesic distance
(\ref{disp-rel-diff}) on the solutions of the geodesic equations (\ref{geodesicEq}) for these metrics.
In doing so one has to be careful to the fact that the metric defining the momentum
space interval (\ref{metricInterval}) has upper indexes $g^{\mu\nu}$,
and the Christoffels (\ref{christoffel}) involve also the inverse
metric $g^{-1}$ since the metric with lower indexes has to be intended
as the one satisfying
\begin{equation}
g_{\mu\rho}g^{\rho\nu}=\delta_{\mu}^{\nu}.
\end{equation}

Starting from the metric (\ref{metricInterval}), one finds
\begin{equation}
\mu^2
= \frac{2}{\ell^2}\left( \cosh\left(\ell m\right) - 1  \right)
= {\cal C}_\lambda(p)
= \left(\frac{2}{\ell}\right)^2 \sinh^{2}\left(\frac{\ell p_{0}}{2}\right)-\vec{p}^{2}e^{-\ell\left(1-2\lambda\right)p_{0}}.
\label{onshellAllOrders}
\end{equation}
Apart from a (trivial) redefinition of the mass, the geodesic distance coincides with the quadratic Casimir (\ref{Casimir}) of $\kappa$-Poincar\'e. We denote $\cal{C}_\lambda$ this mass Casimir, which will play the role of Hamiltonian constraint for the particle action.

The expression of the on-shell relation depends explicitly on the parameter $\lambda$, i.e. it depends on the $\kappa$-Poincaré basis, and coincides with the Casimir~(\ref{Casimir}) in momentum space variables.
The dependence on $\lambda$ remains even at first order in the deformation parameter $\ell$:
\begin{equation}
\mu^2 = m^2 + O(\ell^2)
= p_0^2 - \vec{p}^{2} + \ell\left(1-2\lambda\right)p_{0}\vec{p}^{2}+O(\ell^2).
\end{equation}
This result is in contrast with the one in~\cite{meljanac}, where the form of the on-shell relation (at least at first order in $\ell$) was claimed to be independent on $\lambda$.
One can notice by inspecting the derivation in App.~\ref{sec:on-shell} that the misleading conclusions reached in \cite{meljanac} were mainly due to confusing the role of the metric $g^{\mu\nu}$ and its inverse $g_{\mu\nu}$.

One can see that the on-shell relation for the respective
$\lambda$-parametrized coordinates can be obtained, as discussed earlier,
evaluating the geodesic distance in the Majid-Ruegg coordinates ($\lambda=0$),
and then performing the coordinate change (\ref{mapTorder}) in the expression obtained, corresponding to the (passive) diffeomorphism
\begin{equation}
p_\mu = f_\mu(\tilde{p}) \equiv (\tilde{p}_0, e^{\lambda\ell \tilde{p}_0}\tilde{p}_j).
\label{diffeoTimeOrder}
\end{equation}
This shows, as it should be expected, that in both writings, the physical content of the on-shell relation, i.e. the value of the particle mass, is the same, but the way this
is implemented is by giving an appropriately different form to the on-shell
relation, once the dependence on coordinates is made explicit.

\subsection{``Time-to-the-right'' basis for de Sitter momentum space}
\label{sec:TTR}

We consider first the set-up associated to the so called ``time-to-the-right'' (TTR) basis of $\kappa$-momentum space, which coincides with the ``Majid-Ruegg'' basis discussed in Sec.~\ref{sec:time-order}. This is obtained setting the parameter $\lambda$ characterizing the class of time-ordered bases to zero. From Eq.~(\ref{planeWaves}) we see that this choice corresponds to having ordered the non-commutative time coordinate to the right in the definition of the $\kappa$-Minkowski plane wave (or the $AN(3)$ group elements).
The on-shell relation, sum rule and phase space are thus given by setting $\lambda=0$ in Eqs. (\ref{onshellAllOrders}), (\ref{sumTorder}) and (\ref{phaseSpace}):
\begin{equation}
\mu^2 = {\cal C}(p) = \left(\frac{2}{\ell}\right)^2 \sinh^{2}\left(\frac{\ell p_{0}}{2}\right)-\vec{p}^{2}e^{-\ell p_{0}}.
\label{onshellTTR}
\end{equation}
\begin{equation}
\begin{gathered}
\left( p \oplus q  \right)_0 = p_0 + q_0, \\
\left( p \oplus q  \right)_j = p_j + e^{\ell p_0} q_j.
\end{gathered}
\label{sumTTR}
\end{equation}
\begin{equation}
\begin{gathered}
\left\lbrace x^0,x^j\right\rbrace = \ell x^j, \qquad \left\lbrace x^i,x^j\right\rbrace = 0, \qquad \left\lbrace p_\mu , p_\nu \right\rbrace = 0 , \\
\left\lbrace p_0,x^0 \right\rbrace = 1 , \qquad \left\lbrace p_0,x^j \right\rbrace = 0, \\
\left\{ p_{j},x^{0}\right\} =\ell p_{j}\ ,\qquad\left\{ p_{j},x^{k}\right\} =  \delta_{j}^k,
\label{phaseSpaceTTR}
\end{gathered}
\end{equation}

For the specific process characterized by the conservation law (\ref{conservation}), the  TTR boundary term will have the expression
\begin{equation}
\begin{gathered}
{\cal K}_0 = p_0 \oplus q_0 - k_0 = p_0 +  q_0 - k_j, \\
{\cal K}_j = p_j \oplus q_j - k_j = p_j + e^{\ell p_0} q_j - k_j.
\end{gathered}
\label{boundaryTTR}
\end{equation}
The translations, for the worldlines involved in the process, will be generated by the action by Poisson bracket of ${\cal K}_\mu$.

\subsection{``Time-symmetric'' basis for de Sitter momentum space}
\label{sec:TS}

As a different choice of time-ordered coordinates in momentum space, we consider now the so-called ``time-symmetric'' (TS) basis~\cite{GACagostinikfield2004} $\tilde{p}_\mu$. This is the one obtained choosing the ordering rule for the plane wave~(\ref{planeWaves}) so that the time non-commutative coordinate appears symmetrically (for instance $:\!\!X_i X_0\!\!: \ = \frac{1}{2} (X^0X^i +X^i X^0)$, and amounts to set the parameter $\lambda$ to $\lambda = 1/2$. We get from Eqs. (\ref{onshellAllOrders}), (\ref{sumTorder}) and (\ref{phaseSpace})
\begin{equation}
\mu^2 = \tilde{{\cal C}}(\tilde{p}) = \left(\frac{2}{\ell}\right)^2 \sinh^{2}\left(\frac{\ell \tilde{p}_{0}}{2}\right)-\left( \vec{\tilde{p}} \right)^{2},
\label{onshellTS}
\end{equation}
\begin{equation}
\begin{gathered}
\left( \tilde{p} \tilde{\oplus} \tilde{q}  \right)_0 = \tilde{p}_0 + \tilde{q}_0, \\
\left( \tilde{p} \tilde{\oplus} \tilde{q}  \right)_j = e^{- \frac{1}{2} \ell \tilde{q}_0} \tilde{p}_j + e^{\frac{1}{2} \ell \tilde{p}_0} \tilde{q}_j.
\end{gathered}
\label{sumTS}
\end{equation}
\begin{equation}
\begin{gathered}
\left\lbrace x^0,x^j\right\rbrace = \ell x^j, \qquad \left\lbrace x^i,x^j\right\rbrace = 0, \qquad \left\lbrace \tilde{p}_\mu , \tilde{p}_\nu \right\rbrace = 0 , \\
\left\lbrace \tilde{p}_0, x^0 \right\rbrace = 1 , \qquad \left\lbrace \tilde{p}_0,x^j \right\rbrace = 0, \\
\left\{ \tilde{p}_{j},x^{0}\right\} = \frac{1}{2} \ell \tilde{p}_{j}\ ,\qquad
\left\{ \tilde{p}_{j},x^{k}\right\} = e^{- \frac{1}{2} \ell \tilde{p}_{0}}\delta_{j}^k,
\label{phaseSpaceTS}
\end{gathered}
\end{equation}

For the specific process characterized by the conservation law (\ref{conservation}), the  TTR boundary term will have the expression
\begin{equation}
\begin{gathered}
\tilde{{\cal K}}_0 = \tilde{p}_0 \tilde{\oplus} \tilde{q}_0 - \tilde{k}_0 = \tilde{p}_0 +  \tilde{q}_0 - \tilde{k}_j, \\
\tilde{{\cal K}}_j = \tilde{p}_j \tilde{\oplus} \tilde{q}_j - \tilde{k}_j = e^{- \frac{1}{2} \ell \tilde{q}_0} \tilde{p}_j + e^{\frac{1}{2} \ell \tilde{p}_0} \tilde{q}_j - \tilde{k}_j.
\end{gathered}
\label{boundaryTS}
\end{equation}
The translations, for the worldlines involved in the process, will be generated by the action by Poisson bracket of $\tilde{{\cal K}}_\mu$.

\subsection{Diffeomorphism from TTR to TS momenta}
\label{sec:TTR-TS}

We now consider the change of coordinates relating the TTR and TS $\kappa$-momentum space bases.
From Eq. (\ref{diffeoTimeOrder}) it follows that TTR and TS coordinates are connected through the passive diffeomorphism
\begin{equation}
p_\mu = f_\mu(\tilde{p}) \equiv (\tilde{p}_0, e^{\frac{1}{2}\ell \tilde{p}_0}\tilde{p}_j).
\label{diffeoTTR-TS}
\end{equation}
As shown in Sec. (\ref{sec:on-shell}), and as one can easily verify, under such a diffeomorphism, the on-shell relation (\ref{onshellTTR}) becomes (\ref{onshellTS}).
Consider now the sum law (\ref{sumTTR}). One finds
\begin{equation}
\begin{gathered}
\left( p \oplus q  \right)_0 = p_0 + q_0 = \tilde{p}_0 + \tilde{q}_0 = \left( \tilde{p} \tilde{\oplus} \tilde{q}  \right)_0, \\
\left( p \oplus q  \right)_j = p_j + e^{\ell p_0} q_j = e^{\frac{1}{2}\ell\tilde{p}_0} \tilde{p}_j + e^{\ell \tilde{p}_0} e^{\frac{1}{2} \ell \tilde{q}_0} \tilde{q}_j = e^{\frac{1}{2}\ell(\tilde{p}_0 + \tilde{q}_0)}\left( e^{ - \frac{1}{2} \ell \tilde{q}_0} \tilde{p}_j + e^{\frac{1}{2}\ell \tilde{p}_0} \tilde{q}_j \right)
= e^{\frac{1}{2} \ell (\tilde{p} \tilde{\oplus} \tilde{q})_0} \left( \tilde{p} \tilde{\oplus} \tilde{q}  \right)_j.
\end{gathered}
\label{sumTTR-TS}
\end{equation}
We see that the sum (\ref{sumTTR}) does not changes into (\ref{sumTS}) but in $f_\mu \left( \tilde{p} \tilde{\oplus} \tilde{q}  \right)$, in agreement to the discussion of sec.~\ref{sec:boundary}.
Then the boundary term (\ref{boundaryTTR}) does not change into  (\ref{boundaryTS}), but into
\begin{equation}
\begin{gathered}
\tilde{{\cal K}}^f_0 = f_0 \left( \tilde{p} \tilde{\oplus} \tilde{q} \right) - f_0(\tilde{k}) = \tilde{p}_0 + \tilde{q}_0 - \tilde{k}_0 , \\
\tilde{{\cal K}}^f_j = f_j \left( \tilde{p} \tilde{\oplus} \tilde{q}  \right) - f_j(\tilde{k}) = e^{\frac{1}{2} \ell (\tilde{p}_0 + \tilde{q}_0)} \left( e^{- \frac{1}{2} \ell \tilde{q}_0} \tilde{p}_j + e^{\frac{1}{2} \ell \tilde{p}_0} \tilde{q}_j  \right) - e^{\frac{1}{2} \ell \tilde{k}_0} \tilde{k}_j .
\end{gathered}
\label{boundaryTTR-TS}
\end{equation}
These boundary terms are of the kind discussed in Sec.\ref{sec:boundary}. They produce the same conservation laws of ${\cal K}_\mu$ (\ref{boundaryTS}):
\begin{equation}
 \tilde{{\cal K}}^f_\mu = 0 \quad \Longleftrightarrow \quad
 \left\lbrace
 \begin{split}
  & \tilde{p}_0 + \tilde{q}_0 = \tilde{k}_0, \\
  & e^{- \frac{1}{2} \ell \tilde{q}_0} \tilde{p}_j + e^{\frac{1}{2} \ell \tilde{p}_0} \tilde{q}_j  = \tilde{k}_j ,
 \end{split}
\right.
\qquad \Longleftrightarrow \quad \tilde{{\cal K}}_\mu = 0 \ .
\end{equation}
However $\tilde{{\cal K}}^f_\mu$ and $\tilde{{\cal K}}_\mu$ generate different translations:
\begin{equation}
\begin{gathered}
\left\lbrace {\tilde{\cal K}}^f_0 , \cdot \right\rbrace = \lbrace  \tilde{p}_0 + \tilde{q}_0 - \tilde{k}_0 , \cdot \rbrace , \\
\left\lbrace \tilde{{\cal K}}^f_j , \cdot \right\rbrace =
e^{\frac{1}{2} \ell (\tilde{p}_0 + \tilde{q}_0)} \lbrace \left( \tilde{p} \tilde{\oplus} \tilde{q} \right)_j , \cdot \rbrace
+ \frac{\ell}{2} e^{\frac{1}{2} \ell (\tilde{p}_0 + \tilde{q}_0)} \left( \tilde{p} \tilde{\oplus} \tilde{q} \right)_j \left\lbrace \left( \tilde{p} \tilde{\oplus} \tilde{q} \right)_0 , \cdot \right\rbrace - e^{\frac{1}{2} \ell \tilde{k}_0 } \lbrace \tilde{k}_j, \cdot \rbrace - \frac{\ell}{2} \tilde{k}_j e^{\frac{1}{2} \ell \tilde{k}_0 } \left\lbrace k_0, \cdot \right\rbrace,
\end{gathered}
\end{equation}
which differ from
\begin{equation}
\lbrace \tilde{{\cal K}}_\mu , \cdot \rbrace
= \lbrace  \left( \tilde{p} \tilde{\oplus} \tilde{q} \right)_\mu , \cdot \rbrace - \lbrace \tilde{k}_\mu , \cdot \rbrace .
\end{equation}

\begin{figure}[h!]
\begin{center}
\includegraphics[scale=1]{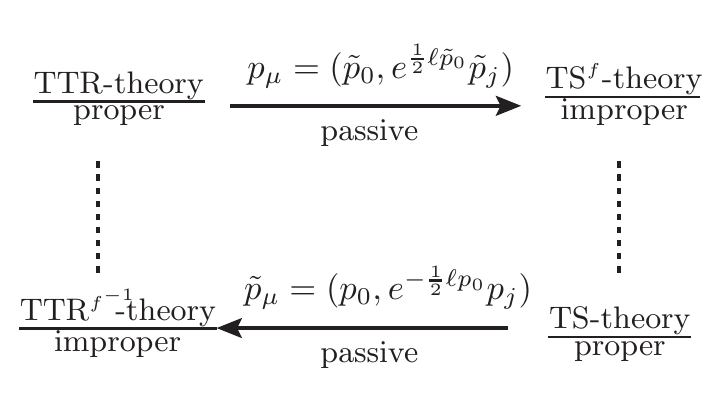}
\end{center}
\caption{Under a passive diffeomorphism from TTR to TS coordinates, the ``proper'' TTR-theory in TTR coordinates is equivalent to an ``improper'' $\text{TS}^f$-theory theory in TS coordinates. Vice versa, when starting from a ``proper'' TS-theory in TS coordinates, which is physically inequivalent to both TTR and TS$'$ theories, the inverse diffeomorphism from TS to TTR coordinates yields an ``improper'' $\text{TTR}^{f^{-1}}$-theory (physically equivalent to the ``proper'' TS-theory).
}
\label{fig:diagramTTR-TS}
\end{figure}

Thus, starting from the ``proper'' theory defined in TTR coordinates (Sec.~\ref{sec:TTR}, let's call it ``TTR-theory'') and performing the change~(\ref{diffeoTTR-TS}) to TS coordinates, does not lead to the theory defined in Sec.~\ref{sec:TS}, characterized by the boundary term $\tilde{{\cal K}}_\mu$, which would be the ``proper'' theory one would write down starting from TS coordinates (let's call it ``TS-theory'').
Instead, the change~(\ref{diffeoTTR-TS}), leads to the ``improper'' theory in TS coordinates, characterized by the boundary term $\tilde{{\cal K}}^f_\mu$, producing the same conservation laws of the TS-theory, but different translation generators (let's call it ``$\text{TS}^f$-theory''). As shown in Fig.~\ref{fig:diagramTTR-TS}, this is the same behaviour depicted in Fig.~\ref{fig:diagramPropImprop} for the general case.

\vspace{-0.5cm}

\newpage

We will see, with an explicit example, how the difference in the translation generators lead, for TS and $\text{TS}^f$ theories, to different predictions for the physical observables. We will show however how, for $\text{TS}^f$-theory and TTR-theory, connected by a passive diffeomorphism~(\ref{diffeoTTR-TS}), the predictions coincide.

\section{Our case study}

\begin{figure}[h!]
\begin{center}
\includegraphics[width=0.5 \textwidth]{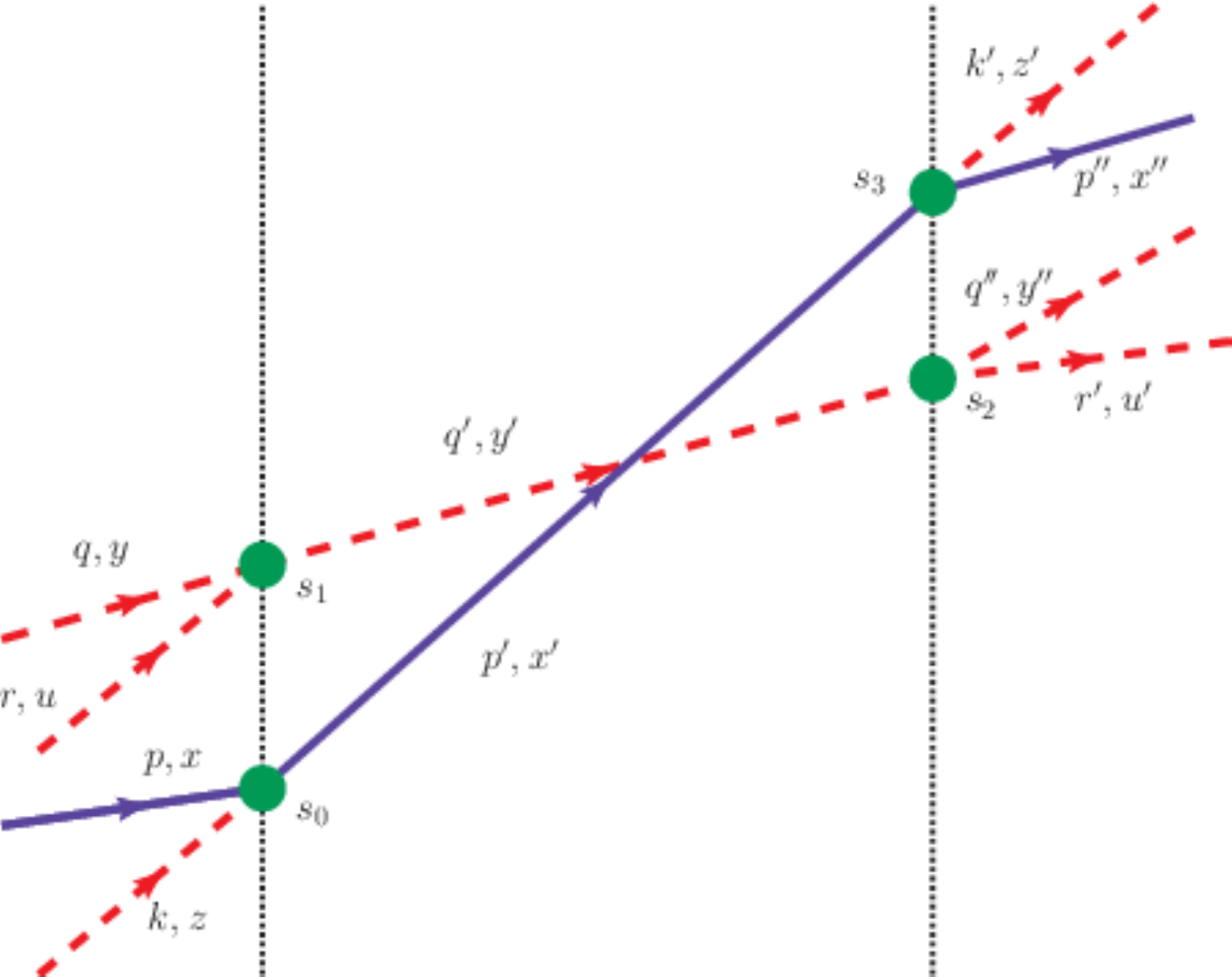}
\caption{
Blue lines represent hard particles, red lines soft particles and black dotted lines stand for the observers.}
\label{fig:digital}
\end{center}
\end{figure}

In this final section we consider an explicit physical example illustrating the features determined in the previous sections. Our analysis will concern the study of a a specific process suitable to compare the time of travels of particles for the TTR, TS, and $\text{TS}^f$ theories respectively.
Let's focus on the specific process depicted in fig.\ref{fig:digital}.
Here we have two atoms ($q,y$ and $p,x$), each one absorbing a photon, propagating freely and finally emitting a second photon.

The energies of the absorbed photons are such that both atoms in the excited states can be considered ultra-relativistic $p'^0 \gg m_{p'}$, $q'^0 \gg m_{q'}$. We work at first order in the deformation ($O(\ell p)$), and we take the case in which $\ell q'^0$ can be neglected,
while $\ell p'^0$ cannot, so that the atom with momentum $q'$ is soft and the one with momentum $p'$ is hard.
We suppose that the soft atom and the hard one are generated at the same spatial point,
but the soft atom is generated after the hard one.
We also suppose that they de-excite at the same spatial point. Then, the main concern of our study will be to predict which of the two atoms reaches the spatial point of de-excitement first.
We remark that the processes involving the soft atom are causally disconnected from the ones involving the hard atom
and this will reflect in the structure of the boundary terms we will propose to describe the processes~\cite{anatomy}.

Before studying the process, let's derive the free particle equations of motion for the generic time-ordered $\kappa$-momentum space.
From the discussions of Secs.~\ref{sec:RelLoc}, \ref{sec:time-order}, and~\ref{sec:on-shell-k-bases}, the Lagrangian of a free particle will be
\begin{equation}
{\cal L}_\lambda(p) = x^\mu \bar{\xi}_\mu^\nu(p) \dot{p}_\nu - {\cal N} \left( {\cal C}_\lambda - \mu^2 \right),
\label{lagrangian}
\end{equation}
with ${\cal C}_\lambda$ given by Eq.~(\ref{onshellAllOrders}).
By variating the Lagrangian in function of $x^a$ and $p_\mu$ one gets the equations of motion
\begin{equation}
\begin{gathered}
\dot{p}_\mu = 0, \quad
{\cal C}_\lambda (p) =\mu^2, \quad \\
\dot{x}^{j}={\cal N}e^{-\lambda\ell p_{0}}\frac{\partial{\cal C}_\lambda\left(p\right)}{\partial p_{j}}, \quad
\dot{x}^{0}={\cal N}\left(\frac{\partial{\cal C}_\lambda\left(p\right)}{\partial p_{0}}+\left(1-\lambda\right)\ell p_{j}\frac{\partial{\cal C}_\lambda\left(p\right)}{\partial p_{j}}\right) ,
\label{eqMotion}
\end{gathered}
\end{equation}
We can solve the equations of motion for $x^j(x^0) = \bar{x}^j + v^j (x^0 - \bar{x}^0)$, evaluating the velocity
\begin{equation}
v^j = \frac{\dot{x}^j}{\dot{x}^0} = \frac{2\ell p_{j}e^{\lambda\ell p_{0}}}{1-e^{2\ell p_{0}}+\ell^{2}\mathbf{p}^{2}e^{2\lambda\ell p_{0}}} ,
\label{velocity}
\end{equation}
where we used the expression (\ref{onshellAllOrders}). Using again Eq. (\ref{onshellAllOrders}) for massless particles, the on-shell relation ${\cal C}_\lambda = 0 $ gives
\begin{equation}
\ell^{2}\mathbf{p}^{2} = \left(e^{(1-\lambda)\ell p_{0}}-e^{-\lambda\ell p_{0}}\right)^{2} ,
\end{equation}
substituting the last expression in (\ref{velocity}) we find, for massless particles,
\begin{equation}
 v^j = - \frac{p_{j}}{|\mathbf{p}|},
 \label{velocityMassless}
\end{equation}
where one should notice that the choice $p_j<0$ ($p_j>0$) coincides, in our covariant conventions, to consider a particle propagating in the direction of the positive (negative) $x^j$ axis. We thus see that, in $x^\mu$ coordinates, the \underline{coordinate} velocity of  massless particles is undeformed. This means, as discussed in previous works~\cite{kappabob,anatomy}, that within this choice of spacetime coordinates, all the features of relative locality manifest themselves in the non trivial role of translation generators.

In the following we work for simplicity in (1+1)D, and at first order in $\ell$, which suffices to illustrate our results.

\subsection{TTR analysis}
\label{sec:TTRanalysis}

We perform first the analysis within the time-to-the-right framework defined in sec.\ref{sec:TTR}, so that for each particle the on-shell relation is given by (\ref{onshellTTR})
and the composition law of momenta is given by (\ref{sumTTR}). Taking into account the expression for the kinetic term (\ref{kineticTerm}) (for $\lambda = 0$), the free-particle part of the action is characterized, for each particle $p^I$, by the free Lagrangian (\ref{lagrangian})
\begin{equation}
 {\cal L}(p^I) = x_I^0 \dot{p}^I_0 + x_I^1 \dot{p}^I_1 - \ell x_I^1 p^I_1 \dot{p}^I_0 + {\cal N}_{I} \left( {\cal C}(p^I) - \mu^2_{I} \right),
 \label{lagrangianTTR}
\end{equation}
where ${\cal N}_p$ is a Lagrange multiplier, enforcing the on-shell condition.
The set of processes in fig.~\ref{fig:digital} is then described by the Relative Locality action
\begin{equation}
\begin{split}
\mathcal{S} = & \int_{-\infty}^{s_{0}} \!\!\!\! ds \left( {\cal L}(k) + {\cal L}(p) \right) +\int_{-\infty}^{s_{1}} \!\!\!\! ds \left( {\cal L}(q) + {\cal L}(r) \right)  +\int_{s_{0}}^{s_{3}} \!\!\!\!  ds {\cal L}(p') + \int_{s_{1}}^{s_{2}} \!\!\!\!  ds {\cal L}(q') + \int_{s_2}^{+\infty} \!\!\!\! ds \left( {\cal L}(q'') + {\cal L}(r') \right) \\
& + \int_{s_3}^{+\infty} \!\!\!\! ds \left( {\cal L}(p'') + {\cal L}(k') \right) -\zeta_{[0]}^{\mu}\mathcal{K}_{\mu}^{[0]}(s_{0})
 -\zeta_{[1]}^{\mu}\mathcal{K}_{\mu}^{[1]}(s_{1})
 -\zeta_{[2]}^{\mu}\mathcal{K}_{\mu}^{[2]}(s_{2})
 -\zeta_{[3]}^{\mu}\mathcal{K}_{\mu}^{[3]}(s_{3})\ ,
\end{split}
\label{actionTTR}
\end{equation}
where the $\zeta_{[i]}$ are Lagrange multipliers enforcing the conservation law at the interaction vertices and play the role of interaction coordinates \cite{RelLocPrinciple, anatomy}. The ${\cal K}^{[i]}_\mu (s_{i})$  are
\begin{equation}
\begin{gathered}
{\cal K}^{[0]}_\mu (s_{0}) = (p\oplus k)_\mu-p'_\mu , \qquad
{\cal K}^{[1]}_\mu (s_{1}) = (q \oplus r)_\mu-q'_\mu  ,\\
{\cal K}^{[2]}_\mu (s_{2}) = q'_\mu-(q'' \oplus r')_\mu , \qquad
{\cal K}^{[3]}_\mu (s_{3}) = p'_\mu - (p''\oplus k')_\mu  .
\label{constraintsTTR}
\end{gathered}
\end{equation}
We notice that in each boundary term one has only the momenta of the particles
which are causally connected with the interaction described by that boundary term. As a result, the action can be split in two parts that do not affect each other:
a part describing the processes concerning the soft atom
and a part describing the processes concerning the hard atom.

From the action~(\ref{actionTTR}), each particle satisfies the equations of motion (\ref{eqMotion}) with $\lambda = 0$, so that each particle moves with velocity $v_I^1 = \pm 1$. Moreover the boundary terms in (\ref{actionTTR}) enforce the conservation laws at the vertices ${\cal K}^{[i]}_\mu (s_{i})=0$.
From the variation of the action (\ref{actionTTR}) one obtains the boundary conditions
\begin{equation}
 x_I^\mu(s_{i}) = \pm \zeta^\nu_{[i]} \left(\frac{\partial \mathcal{K}^{[i]}_\nu}{\partial p^I_\mu}
 + \ell \delta^\mu_0 \frac{\partial \mathcal{K}^{[i]}_\nu}{\partial p^I_1}p^I_1\right) ,
\label{boundaryCondTTR}
\end{equation}
where the $+$/$-$ sign is for particles incoming/outgoing at the vertex.
Notice that the boundary conditions~(\ref{boundaryCondTTR}) can be also expressed in terms of Poisson brackets as explained in App.~\ref{app:translationsGeneric}, through Eq.~(\ref{boundaryCondPB}).

To determine the time of arrival of the two atoms we introduce an observer Alice which is local to the excitation of the atoms and an observer Bob which is local to their de-excitation.
We take these observers to be in relative rest, so that the relation between their coordinates is given by a translation transformation.
Considering that (taking $p'_1,q'_1<0$ in order to have propagation in the positive direction of the $x^1$ axis) the atoms move with velocity (\ref{velocityMassless}) $v^1 = 1$, the relativistic properties of the theory ensures that each observer describes the particle worldlines
\begin{equation}
x_I^1 = \bar{x}_I^1 \pm (x_I^0 - \bar{x}_I^0).
\label{eqMotionO}
\end{equation}
Let us suppose that the hard atom is generated at Alice's coordinates $x'^1_A=x'^0_A=0$, while the soft atom at $y'^1_A=0$, $y'^0_A=t_0>0$.
Taking $p'_1,q'_1<0$ to have propagation in the positive direction of the $x^1$ axis, it follows that Alice's worldlines are
\begin{equation}
 {x'}_{A}^{1} ={x'}_{A}^{0} , \qquad \qquad
 {y'}_{A}^{1} ={y'}_{A}^{0}- t_0 .
 \label{eq-motion-alice}
\end{equation}
We suppose that the de-excitation of the two atoms occurs at Bob spatial origin $x'^1_B=y'^1_B=0$.
To compute the times at which these events happen, we use the worldlines described by Bob.
These worldlines can be obtained by introducing in  (\ref{eqMotionO}) the translation transformation which relates the coordinates of Alice and Bob.

As discussed above, it follows from the action (\ref{actionTTR}) that assuming, according to~\cite{anatomy},  the interaction coordinates to translate as $\zeta_{[i]B}^\mu = \zeta_{[i]A}^\mu + b^\mu$ (see App.~\ref{app:translations}), the translations are generated by the Poisson brackets of the boundary terms ${\cal K}^{[i]}_\mu$ with the coordinates (with the appropriate sign for incoming and outgoing particles), so that
\begin{equation}
\begin{split}
 {x'}_{B}^{\mu}(s) & = {x'}_{A}^{\mu}(s) + b^{\nu}  \{ {\cal K}^{[0]}_\nu , x'^\mu\} = {x'}_{A}^{\mu}(s) - b^{\nu}  \{ {\cal K}^{[3]}_\nu , x'^\mu\}
 = {x'}_{A}^{\mu}(s) - b^{\nu}  \{ p'_\nu , x'^\mu\}\ , \\
 {y'}_{B}^{\mu}(s)& = {y'}_{A}^{\mu}(s) + b^{\nu}  \{ {\cal K}^{[1]}_\nu , y'^\mu\}
 = {y'}_{A}^{\mu}(s) - b^{\nu}  \{ {\cal K}^{[2]}_\nu , y'^\mu\}
  ={y'}_{A}^{\mu}(s) - b^{\nu}  \{ q'_\nu , y'^\mu\}\ ,
\end{split}
\label{translationsTTR}
 \end{equation}
which, written explicitly using (\ref{phaseSpaceTTR}), give
\begin{equation}
\begin{split}
& {x'}_{B}^{0}(s) ={x'}_{A}^{0}(s)-b^0-b^1\ell p'_1 \ , \qquad \qquad \qquad ~~~~~~
 {x'}_{B}^{1}(s) ={x'}_{A}^{1}(s)-b^1 \ , \\
& {y'}_{B}^{0}(s) ={y'}_{A}^{0}(s)-b^0-b^1\ell q'_1 \simeq {y'}_{A}^{0}(s)-b^0 \ , \qquad
 {y'}_{B}^{1}(s) ={y'}_{A}^{1}(s)-b^1 \ ,
 \label{translation-hardTTR}
\end{split}
\end{equation}
where we have neglected $b^1\ell q'_1$.
Substituting these relations for $\bar{x}_B^\mu(\bar{x}_A)$ in (\ref{eqMotionO}), we find that Bob describes the worldlines
 \begin{equation}
   x'^1_B =x'^0_B  -b^1+b^0+b^1\ell p'_1\ , \qquad \qquad
   y'^1_B =y'^0_B -b^1+b^0-t_0\ .
   \label{eq-motion-bobTTR}
 \end{equation}
When we impose $x'^1_B=y'^1_B=0$, we find:
 \begin{equation}
  \begin{split}
   x'^0_B&=+b^1-b^0 + |b^1\ell p'_1|\ , \\
   y'^0_B&=+b^1-b^0+t_0\ ,
  \end{split}
 \end{equation}
where we used that $-b^1\ell p'_1 = |b^1\ell p'_1|$.
So, if  $t_0 <  |b^1\ell p'_1|$, the soft atom arrives at Bob before the hard atom, even if it was emitted later at Alice.

\subsection{TS analysis}
\label{sec:TSanalysis}

A similar analysis can be carried out in the time-symmetric set-up defined in sec.\ref{sec:TS}.
The on-shell relation for every particle is now given by (\ref{onshellTS}), while the  composition law of momenta by (\ref{sumTS}).
The kinetic term is (\ref{kineticTerm}) with $\lambda = 1/2$, so that the free-particle part of the action is characterized by the free Lagrangian (\ref{lagrangian})
\begin{equation}
\tilde{{\cal L}}(\tilde{p}^I) = x^0 \dot{\tilde{p}}_0 + x^j \dot{\tilde{p}}_j - \frac{1}{2} \ell x^1 \tilde{p}_1 \dot{\tilde{p}}_0  + \frac{1}{2} x^1 \ell \tilde{p}_0 \dot{\tilde{p}}_1  +  {\cal N}_I \left( \tilde{{\cal C}}(\tilde{p}^I) - \mu^2 \right).
\label{lagrangianTS}
\end{equation}
The physical configuration of fig.~\ref{fig:digital} is again described by the action (\ref{actionTTR}) where one has to substitute ${\cal L}(p^I) \rightarrow \tilde{{\cal L}}(\tilde{p}^I)$ and ${\cal K}^{[i]}_\mu \rightarrow \tilde{{\cal K}}^{[i]}_\mu $, with $\tilde{{\cal K}}^{[i]}_\mu$ of the kind (\ref{boundaryTS}), i.e. they are (see (\ref{constraintsTTR}))
\begin{equation}
\begin{gathered}
\tilde{{\cal K}}^{[0]}_\mu (s_{0}) = (\tilde{p}\tilde{\oplus} \tilde{k})_\mu-\tilde{p}'_\mu , \qquad
\tilde{{\cal K}}^{[1]}_\mu (s_{1}) = (\tilde{q} \tilde{\oplus} \tilde{r})_\mu-\tilde{q}'_\mu  ,\\
\tilde{{\cal K}}^{[2]}_\mu (s_{2}) = \tilde{q}'_\mu-(\tilde{q}'' \tilde{\oplus} \tilde{r}')_\mu , \qquad
\tilde{{\cal K}}^{[3]}_\mu (s_{3}) = \tilde{p}'_\mu - (\tilde{p}''\tilde{\oplus} \tilde{k}')_\mu  .
\label{constraintsTS}
\end{gathered}
\end{equation}

Each particle satisfies now the equations of motion (\ref{eqMotion}) with $\lambda = 1/2$, and from (\ref{velocityMassless}) it follows that each particles moves again with velocity $v_I^1 = \pm 1$.
The boundary conditions following from the variation of the TS action are different from (\ref{boundaryCondTTR}) due to the difference in the kinetic term in (\ref{lagrangianTS}). They are
\begin{equation}
\tilde{x}_I^{\mu}(s_i)=\pm\zeta_{\left[i\right]}^{\nu} \left(  \frac{\partial\tilde{{\cal K}}_{\nu}^{\left[i\right]}}{\partial\tilde{p}^I_{\mu}}
+ \frac{1}{2} \ell \left( \delta^\mu_0 \tilde{p}^I_1
- \delta_{1}^{\mu} \tilde{p}^I_{0}  \right) \frac{\partial\tilde{{\cal K}}_{\nu}^{\left[i\right]}}{\partial\tilde{p}^I_{1}} \right) .
\label{boundaryCondTS}
\end{equation}

Repeating the same steps as for the TTR case, noticing that since $v^1 = \pm 1$ Eq. (\ref{eqMotionO}) and (\ref{eq-motion-alice}) still hold, and also that one can show that Eqs.~(\ref{translationsTTR}), replacing ${\cal K}_\mu $ with $\tilde{{\cal K}}_\mu$ and $p_\mu$ with $\tilde{p}_\mu$, hold for the TS case, we find, using (\ref{phaseSpaceTS}), that (\ref{translation-hardTTR}) are replaced by
\begin{equation}
\begin{split}
& {x'}_{B}^{0}(s) ={x'}_{A}^{0}(s)-b^0- \frac{1}{2} b^1\ell \tilde{p}'_1 \ , \qquad \qquad \qquad ~~~~~~
 {x'}_{B}^{1}(s) ={x'}_{A}^{1}(s)-b^1 + \frac{1}{2} b^1\ell \tilde{p}'_0\ , \\
& {y'}_{B}^{0}(s) ={y'}_{A}^{0}(s)-b^0- \frac{1}{2} b^1\ell \tilde{q}'_1 \simeq {y'}_{A}^{0}(s)-b^0 \ , \qquad
 {y'}_{B}^{1}(s) ={y'}_{A}^{1}(s)-b^1 + \frac{1}{2} b^1\ell \tilde{q}'_0 \simeq {y'}_{A}^{1}(s)-b^1\ ,
 \label{translation-hardTS}
\end{split}
\end{equation}
Substituting these relations for $\bar{x}_B^\mu(\bar{x}_A)$, we find that Bob describes the worldlines
 \begin{equation}
    x'^1_B =x'^0_B  -b^1+b^0+ \frac{1}{2}b^1 \ell \left( \tilde{p}'_0 + \tilde{p}'_1 \right)\ , \qquad \qquad
   y'^1_B =y'^0_B -b^1+b^0-t_0\ .
   \label{eq-motion-bobTS}
 \end{equation}
Notice now that on-shell $\tilde{p}'_0 = |\tilde{p}'_1 | + O(\ell (\tilde{p}'_1)^2) = - \tilde{p}'_1 + O(\ell (\tilde{p}'_1)^2) $, so that, imposing $x'^1_B=y'^1_B=0$, we find:
 \begin{equation}
  \begin{split}
   x'^0_B&=+b^1-b^0 \ , \\
   y'^0_B&=+b^1-b^0+t_0\ .
  \end{split}
 \end{equation}
In the TS framework introduced in sec.\ref{sec:TS}, contrary to the TTR framework, the soft atom arrives at Bob always after the hard photon.

\subsection{Diffeomorphism from TTR to TS momenta}

We consider here the same process, fig.~\ref{fig:digital}, analyzed in the framework discussed in sec.~\ref{sec:TTR-TS}, i.e. the one obtained from the TTR framework performing a (passive) diffeomorphism (\ref{diffeoTTR-TS}) on the momentum space coordinates, changing from TTR to TS coordinates:
\begin{equation}
p_\mu = f_\mu(\tilde{p}) \equiv (\tilde{p}_0, e^{\frac{1}{2}\ell \tilde{p}_0}\tilde{p}_j).
\label{diffeoTTR-TS-2}
\end{equation}
One can easily verify that under the change (\ref{diffeoTTR-TS-2}) the free particle Lagrangian (\ref{lagrangianTTR}) is mapped into the Lagrangian (\ref{lagrangianTS}):
\begin{equation}
{\cal L}(p^I) = {\cal L}(f(p^I)) = \tilde{{\cal L}}(\tilde{p}^I) .
\end{equation}
Thus the free part of the action is the same of the TS one of sec.\ref{sec:TSanalysis}, and generates the same equations of motion.
The only difference with the TS action is in the interaction boundary terms, which are of the kind $\tilde{{\cal K}}^f_\mu$ (\ref{boundaryTTR-TS}), i.e. respect to~(\ref{constraintsTS}), one has to substitute to each sum $(\tilde{p}\tilde{\oplus}\tilde{q})_\mu$ the one (see (\ref{sumTTR-TS})) obtained by the diffeomorphism $f_\mu(\tilde{p}\tilde{\oplus}\tilde{q})$.
For instance the boundary terms at $s_0$ and $s_1$ change into
\begin{equation}
\begin{gathered}
 \tilde{{\cal K}}^{f[0]}_0 = (\tilde{p}\tilde{\oplus} \tilde{k})_0-\tilde{p}'_0 , \qquad
 \tilde{{\cal K}}^{f[0]}_1 = e^{\frac{1}{2}\ell (p_0 + k_0)}(\tilde{p}\tilde{\oplus} \tilde{k})_1 - e^{\frac{1}{2}\ell p'_0} \tilde{p}'_1 ,\\
 \tilde{{\cal K}}^{f[0]}_0 = (\tilde{q}\tilde{\oplus} \tilde{r})_0-\tilde{q}'_0 , \qquad
 \tilde{{\cal K}}^{f[0]}_1 = e^{\frac{1}{2}\ell (q_0 + r_0)}(\tilde{q}\tilde{\oplus} \tilde{r})_1 - e^{\frac{1}{2}\ell q'_0} \tilde{p}'_1 ,
\end{gathered}
\label{boundaryCondTTR-TS}
\end{equation}
The boundary conditions are obtained by replacing $\tilde{{\cal K}}^{[i]}_\mu$ with $\tilde{{\cal K}}^{f[i]}_\mu$ in~(\ref{boundaryCondTS}), and with the same mechanism shown in App.~\ref{app:translations} one finds that the translations are generated by the action by Poisson brackets of $\tilde{{\cal K}}^{f[i]}_\mu$.
Then, using (\ref{phaseSpaceTS}) together with (\ref{boundaryCondTTR-TS}), Bob's coordinates are
\begin{equation}
\begin{split}
& {x'}_{B}^{0}(s) ={x'}_{A}^{0}(s)-b^0- b^1\ell \tilde{p}'_1 \ , \qquad \qquad \qquad ~~~~~~
 {x'}_{B}^{1}(s) ={x'}_{A}^{1}(s)-b^1 \ , \\
& {y'}_{B}^{0}(s) ={y'}_{A}^{0}(s)-b^0 -   b^1\ell \tilde{q}'_1 \simeq {y'}_{A}^{0}(s)-b^0 \ , \qquad
 {y'}_{B}^{1}(s) ={y'}_{A}^{1}(s)-b^1 \ ,
 \label{translation-hardTTR-TS}
\end{split}
\end{equation}
Substituting these relations for $\bar{x}_B^\mu(\bar{x}_A)$, we find that Bob describes the worldlines
\begin{equation}
   x'^1_B =x'^0_B  -b^1 + b^0 +  b^1 \ell \tilde{p}'_1 \ , \qquad \qquad
   y'^1_B =y'^0_B -b^1+b^0-t_0\ ,
\end{equation}
so that, imposing $x'^1_B=y'^1_B=0$, we find that Bob detects the particles at the times
 \begin{equation}
  \begin{split}
   x'^0_B&=+b^1-b^0 +  |b^1 \ell \tilde{p}'_1|\ , \\
   y'^0_B&=+b^1-b^0+t_0\ .
  \end{split}
 \end{equation}
The difference in the arrival time of the two photons is $|b^1 \ell \tilde{p}'_1| - t_0 = |b^1 \ell p'_1| - t_0  $, as for the TTR case of sec.\ref{sec:TTRanalysis}.

Thus we see that, while the observable arrival-time derived in the ``proper'' TTR framework of Secs.~\ref{sec:TTR},~\ref{sec:TTRanalysis} differs with the one derived in the ``proper'' TS framework of Secs.~\ref{sec:TS},~\ref{sec:TSanalysis}, it coincides with the ``improper'' framework $\text{TS}^f$ in TS coordinates, obtained from the TTR by a (passive) diffeomorphism.
Thus, this example shows explicitly the meaning of the diagram depicted in Fig.~\ref{fig:diagramTTR-TS} (or in Fig.~\ref{fig:diagramPropImprop}): while passive diffeomorphisms connect a ``proper'' theory with a physically equivalent improper theory, ``proper'' theories in different momentum space coordinate bases yields different physical predictions, as they are not connected by passive diffeomorphisms.

\subsection{Generic (passive) diffeomorphism}

The result obtained in the last subsection is valid for a generic diffeomorphism. Indeed consider a diffeomorphism
\begin{equation}
 \label{reparametrization}
p_\mu=f_\mu(\tilde{p}_\nu) .
\end{equation}
The free Lagrangian changes into ${\cal L}^f(\tilde{p}) = {\cal L}(f(\tilde{p}))$. The variation of ${\cal L}^f(\tilde{p})$ respect to $x^a$ will give the equations of motion
\begin{equation}
\frac{\partial f_\mu (\tilde{p})}{\partial \tilde{p}_\nu} \dot{\tilde{p}}_\nu = 0 \qquad \Longleftrightarrow \qquad \dot{\tilde{p}}_\nu = 0.
\label{mom-conserved}
\end{equation}
The variation respect to $\tilde{p}_\mu$ gives the equations of motion
\begin{equation}
\begin{split}
& \frac{\partial {\cal L}^f (\tilde{p})}{\partial \tilde{p}_\mu} - \frac{d}{ds}  \frac{\partial {\cal L}^f (\tilde{p})}{\partial \dot{\tilde{p}}_\mu}
= \frac{\partial {\cal L} (f(\tilde{p}))}{\partial f_\nu} \frac{\partial f_\nu(\tilde{p})}{\partial \tilde{p}_\mu} - \frac{d}{ds}  \frac{\partial {\cal L} (f(\tilde{p}))}{\partial \dot{f}_\nu} \frac{\partial \dot{f}_\nu(\tilde{p})}{\partial \dot{\tilde{p}}_\mu} \\
& = \frac{\partial f_\nu(\tilde{p})}{\partial \tilde{p}_\mu} \left(  \frac{\partial {\cal L} (p)}{\partial p_\nu}
- \frac{d}{ds}  \frac{\partial {\cal L} (p)}{\partial \dot{p}_\nu} \right)\Big|_{p=f(\tilde{p})} = 0,
\end{split}
\label{eqMotionDiffeo}
\end{equation}
where we used that $\partial {\cal L}(f(\tilde{p}))/\partial f_\mu = \partial {\cal L}(p)/\partial p_\mu $, that $\partial \dot{f}_\nu(\tilde{p}) / \partial \dot{\tilde{p}}_\mu = \partial f_\nu(\tilde{p}) / \partial \tilde{p}_\mu $ and (\ref{mom-conserved}). Multiplying the last expression by $(\partial f_\nu / \partial \tilde{p}_\mu)^{-1}$, it follows that the equations of motion are the same of the  original action, just rewritten in the new coordinate system.
Similarly, as it is shown in appendix \ref{app:translationDiffeo} and in particular Eq.~(\ref{translationDiffeoPB}), the boundary terms generated by the $\tilde{p}_\mu$ variation couple with the interaction boundary terms ${\cal K}^f_\mu$ of the kind (\ref{boundaryTTR-TS}) in such a way that the translations are generated by Poisson brackets with ${\cal K}^f_\mu$:
\begin{equation}
x_{B}^\mu = x_{A}^\mu \pm b^\nu \{ {{\cal K}^f}^{[i]}_\nu (\tilde{p}), x^\mu\}.
\end{equation}
From Eq.~(\ref{translationDiffeoPB-repar}) this implies that translations are obtained by simply rewriting the translations for the original action in terms of the new momenta variables:
\begin{equation}
 x_B^\mu = x_A^\mu \pm b^{\nu}  \{ {\cal K}_{\nu}(p) , x^\mu \}\Big|_{p=f(\tilde{p})} .
\end{equation}
This, together with (\ref{eqMotionDiffeo}), allows us to write the analysis in the framework obtained by diffeomorphism from the TTR one of \ref{sec:TTRanalysis} by simply rewriting the equations of motion for Alice and Bob of \ref{sec:TTRanalysis} in terms of the new variables. In particular Alice still describes the equations (\ref{eq-motion-alice}):
\begin{equation}
 {x'}_{A}^{1} ={x'}_{A}^{0} , \qquad \qquad
 {y'}_{A}^{1} ={y'}_{A}^{0}- t_0 .
 \label{eq-motion-alice-reparametrized}
\end{equation}
The translation transformations for the coordinate of the soft and hard atoms are:
\begin{equation}
\begin{gathered}
 x'^{0}_{B}(s) =x'^{0}_{A}(s)-b^0-b^1\ell f_1(\tilde{p}') \ , \qquad \qquad
 x'^{1}_{B}(s) =x'^{1}_{A}(s)-b^1 \ , \\
 y'^{0}_{B}(s) =y'^{0}_{A}(s)-b^0-b^1\ell f_1(\tilde{q}') \simeq y'^{0}_{A}(s)-b^0 \ , \qquad \qquad
 y'^{1}_{B}(s) =y'^{1}_{A}(s)-b^1 \ .
 \label{translation-hard-reparametrized}
\end{gathered}
\end{equation}
So, putting together (\ref{eq-motion-alice-reparametrized}) and (\ref{translation-hard-reparametrized}), and using
that the de-excitation of the two atoms still occurs at Bob spatial origin $x'^1_B=y'^1_B=0$, we find
 \begin{equation}
  \begin{split}
   x'^0_B&=+b^1-b^0 +|b^1\ell f_1(p')|\ , \\
   y'^0_B&=+b^1-b^0+t_0\ ,
  \end{split}
 \end{equation}
 which are the same results found in the previous section, just rewritten in terms of the new coordinates.
Then also the condition for the soft atom to arrive at Bob before the hard atom is simply the one of the previous section
rewritten in the new coordinate system:
\begin{equation}
t_0< |b^1\ell f_1(p')| \ .
\end{equation}
We can conclude that, after a change of coordinates, the predictions of the theory do not change, but they get just rewritten in the new coordinate system.

\section{Conclusions and comments}
\label{sec:conclusions}

In this paper we studied the effect of a change of momentum-space coordinates when momentum space  is curved, relying on the relative locality framework proposed in~\cite{RelLocPrinciple} and further developed in several following works (see, {\it e.g.} Refs.~\cite{FreSmoGRBrelLoc,anatomy,CarmonaRelLoc,FlaGiukRelLoc,spinning,FreidelSnyderRelLoc,JurekReview,causality,palmisano,MigSamSnyderRL,MigRosSnyderRelLoc,multipart,CarmonaLocal,GiuliaBoosts}).
While most of our findings apply in general to the relative-locality framework~\cite{RelLocPrinciple}, we mainly focused on
DSR-relativistic pictures, where the entire set of (deformed) relativistic spacetime symmetries is available (and momentum space is maximally symmetric).

We found, as expected, that the on-shell relation, defined by the geodesic length in momentum-space, in its explicit form, depends on the (momentum-space) coordinates just in the way needed for its value to be invariant. This also rectifies some erroneous conclusions drawn in~\cite{meljanac}.
For what concerns the boundary terms we considered two different choices of coordinates, $p_\mu$ and $\tilde{p}_\mu$,
 and found that the respective actions, $S(p_\mu)$ and $S(\tilde{p}_\mu)$, lead in general to different predictions for physical observables. We have shown that a diffeomorphism in momentum space $p_\mu = f_\mu(\tilde{p}_\mu)$ does not map the action $S(p_\mu)$ to the action $S(\tilde{p}_\mu)$, but it maps it into an action $S^f(\tilde{p}_\mu)$ which has the same equations of motion and conservation laws of $S(\tilde{p}_\mu)$, but different boundary terms, which in turn generate different translational symmetries with respect to $S(\tilde{p}_\mu)$. The difference in translational symmetries is such that the observables of the theory $S^f(\tilde{p}_\mu)$ coincide with the ones of the theory $S(p_\mu)$.
Indeed, the result of the diffeomorphism amounts to a reparametrization of the boundary terms, originally written in coordinates $p_\mu$, in terms of the new coordinates $\tilde{p}_\mu$, under which the physical predictions do not change, while in general, different boundary terms, characterizing different translation generators, define different theories, even if they encode the same conservation law.

Specifically for the kappa-momentum space, which was our case study, we found that different $\kappa$-Poincar\'e bases lead to inequivalent relativistic theories.
We obtained this result relying on a specific example of physical observable, which we identified with the temporal ordering (time delay) between two events corresponding to the de-excitation of two freely propagating atoms of different energies at a common spatial point, measured by observers local to the two events. 
We focused on the study of time-delays both because of its conceptual interest, and because of its relevance for quantum gravity phenomenology~\cite{GACMavroGRBs,IceCubeNature}.
Our conclusion is that at least for the example we considered, the physical content of the theory depends on the momentum space basis (in the sense explained above).
It would be of great interest to test our results considering alternative observables other than time-delays.

Our results rely moreover on a certain set of prescriptions for how to write the relative locality action.
For instance, we have assumed that 
the propagation of the two atoms, which are taken to be not ``causally connected'', can be described by two separate (independent) relative locality actions. 
This assumption is coherent with a commonly adopted perspective~\cite{RelLocPrinciple,causality,multipart} for which the total action of two chains of processes which are not causally connected is made of the ordinary sum of two relative locality actions.
However, it is still legitimate and interesting to study possible alternatives, as for instance the case in which the total action cannot be decomposed into the sum of two actions each one depending on a different set of variables.
Another prescription, as stressed in Sec.~\ref{sec:boundary}, regards the form of the interaction boundary term to be taken to construct the relative locality action starting from a given set of momentum space coordinates. We have assumed that the boundary term should be given by the ordinary difference of the ``total'' incoming and outgoing momenta, which has proven to ensure compatibility with translational symmetry~\cite{anatomy,GiuliaBoosts}.

We conclude that, within the given set of prescriptions,  under a generic (passive) diffeomorphism in momentum space, a theory with curved momentum space yields the same physical predictions, but its  action is not invariant in form.
This spoils the equivalence between passive and active diffeomorphisms (in momentum space), while in general relativity, on the other hand, such equivalence for spacetime diffeomorphisms is at the basis of general covariance.

A further comment should be made for how, in the relative locality framework, diffeomorphisms affect a theory invariant under ordinary special relativistic (Poincar\'e) symmetries.
The special relativistic limit must be indeed obtained from the relative locality action in the limit of vanishing momentum space curvature, i.e. when momentum space is (Minkowskian) flat.
The behaviour under diffeomorphisms of a theory with flat momentum space is essentially the same as the behaviour, that we have studied in this paper, of a theory with relativistic curved momentum space:
under a passive diffeomorphism, corresponding to a generic change of momentum space coordinates, the physical content of the theory is unchanged; however, the action will be not invariant in form~\footnote{The theory is invariant only under the subsector of diffeomorphisms corresponding to Poincar\'e transformations for flat momentum space, or to the respective relativistic transformations for de Sitter momentum space
(for instance $\kappa$-Poincar\'e symmetries for $\kappa$-momentum space).}. 
In other words, the theory is (obviously) invariant under a passive diffeomorphism, which corresponds simply to a relabeling of momenta, but it is not invariant under the corresponding active diffeomorphism.

However, while our result holds identically for flat and curved momentum spaces, its significance is greater in the curved-momentum-space case.
This is because on a curved geometry (even a maximally-symmetric one) there is no ``natural''  choice of coordinates for constructing the action, while for a flat geometry one has such a natural coordinatization.
For instance, in $\kappa$-(de Sitter-)momentum space one can construct the action starting from a class of possible coordinates corresponding to different bases of $\kappa$-Poincar\'e, leading to relativistic theories that are equally legitimate from the theoretical point of view, but 
are connected by active diffeomorphisms and, as we have shown in this paper, yield different predictions for the observation of time-delays.
It follows that theories formulated in different bases of the same curved momentum space are in general inequivalent, and lead to different physical predictions, that need to be studied experimentally.

While our whole analysis relied on the relative-locality framework, we expect that similar results will be found
in any formalism allowing for momentum-space curvature. Indeed, our findings suggest that 
the duality between spacetime and momentum-space, while geometrically appealing when both are described as curved manifolds, physically is not fully realized, a feature which in the relative-locality framework becomes evident upon observing that when one take into account
the boundary terms, which constrain the motion of particles to be physical,  the different nature of the two spaces becomes relevant.
We also observe that the invariance of spacetime under diffeomorphisms, strictly related to the notion of general covariance, can be motivated by the presence of the gravitational field, while
an analogous physical source for momentum-space curvature as so far not been discussed in the literature and is not expected.

\section*{Acknowledgements}

We are grateful to Riccardo J. Buonocore for contributing to the initial stages of this project.
We thank the referee for the valuable comments.

\appendix

\section{On-shell relation for ``time-ordered'' parametrized momentum space
metrics}
\label{sec:on-shell}

In this section we show explicitly how to obtain the on-shell relation~(\ref{onshellAllOrders}) from the metric~(\ref{metricInterval}).
To find the on-shell relation we have to evaluate the geodesic distance
(\ref{disp-rel-diff}) on the solutions of the geodesic equations (\ref{geodesicEq}).
One has to be careful to the fact that the metric defining the momentum
space interval (\ref{metricInterval}) has upper indexes $g^{\mu\nu}$,
and the Christoffels (\ref{christoffel}) involve also the inverse
metric $g^{-1}$.
One gets, from (\ref{metricInterval}),
\begin{equation}
g^{\mu\nu}\equiv g\left(\gamma\right)=\left(\begin{array}{cccc}
1-\ell^{2}\lambda^{2}e^{-2\ell\left(1-\lambda\right)\gamma_{0}}\vec{\gamma}^{2} & -\ell\lambda e^{-2\ell\left(1-\lambda\right)\gamma_{0}}\gamma_{1} & -\ell\lambda e^{-2\ell\left(1-\lambda\right)\gamma_{0}}\gamma_{2} & -\ell\lambda e^{-2\ell\left(1-\lambda\right)\gamma_{0}}\gamma_{3}\\
-\ell\lambda e^{-2\ell\left(1-\lambda\right)\gamma_{0}}\gamma_{1} & -e^{-2\ell\left(1-\lambda\right)\gamma_{0}} & 0 & 0\\
-\ell\lambda e^{-2\ell\left(1-\lambda\right)\gamma_{0}}\gamma_{2} & 0 & -e^{-2\ell\left(1-\lambda\right)\gamma_{0}} & 0\\
-\ell\lambda e^{-2\ell\left(1-\lambda\right)\gamma_{0}}\gamma_{3} & 0 & 0 & -e^{-2\ell\left(1-\lambda\right)\gamma_{0}}
\end{array}\right).\label{metric}
\end{equation}
The inverse metric is
\begin{equation}
g_{\mu\nu}\equiv g^{-1}\left(\gamma\right)=\left(\begin{array}{cccc}
1 & -\ell\lambda\gamma_{1} & -\ell\lambda\gamma_{2} & -\ell\lambda\gamma_{3}\\
-\ell\lambda\gamma_{1} & \ell^{2}\lambda^{2}\gamma_{1}^{2}-e^{2\ell\left(1-\lambda\right)\gamma_{0}} & \ell^{2}\lambda^{2}\gamma_{1}\gamma_{2} & \ell^{2}\lambda^{2}\gamma_{1}\gamma_{3}\\
-\ell\lambda\gamma_{2} & \ell^{2}\lambda^{2}\gamma_{1}\gamma_{2} & \ell^{2}\lambda^{2}\gamma_{2}^{2}-e^{2\ell\left(1-\lambda\right)\gamma_{0}} & \ell^{2}\lambda^{2}\gamma_{2}\gamma_{3}\\
-\ell\lambda\gamma_{3} & \gamma_{1}\gamma_{3}\left(-\ell\right)^{2}\lambda^{2} & \ell^{2}\lambda^{2}\gamma_{2}\gamma_{3} & \ell^{2}\lambda^{2}\gamma_{3}^{2}-e^{2\ell\left(1-\lambda\right)\gamma_{0}}
\end{array}\right).\label{metricINV}
\end{equation}
From (\ref{metric}) and (\ref{metricINV}) one gets the Christoffels
\begin{equation}
\begin{gathered}\Gamma_{0}^{\ 00}=-\lambda^{2}\ell^{3}\vec{\gamma}^{2}e^{-2\ell\left(1-\lambda\right)\gamma_{0}},\\
\Gamma_{0}^{\ 0j}=\Gamma_{0}^{\ j0}=-\ell^{2}\gamma_{j}e^{-2\ell\left(1-\lambda\right)\gamma_{0}},\\
\Gamma_{0}^{\ jk}=-\ell e^{-2\ell\left(1-\lambda\right)\gamma_{0}}\delta_{jk},\\
\Gamma_{j}^{\ 00}=-\lambda\ell^{2}\gamma_{j}\left(2-\lambda-\lambda^{2}\ell^{2}\vec{\gamma}^{2}e^{-2\ell\left(1-\lambda\right)\gamma_{0}}\right),\\
\Gamma_{j}^{\ 0k}=\Gamma_{j}^{\ k0}=\lambda^{2}\ell^{3}\gamma_{j}\gamma_{k}e^{-2\ell\left(1-\lambda\right)\gamma_{0}},\\
\Gamma_{j}^{\ kl}=\lambda\ell^{2}\gamma_{j}e^{-2\ell\left(1-\lambda\right)\gamma_{0}}\delta_{kl}.
\end{gathered}
\end{equation}
One has to substitute them in Eq.~(\ref{geodesicEq}) and solve
for $\gamma\left(s\right)$. Let's restrict to the 1+1D case for simplicity,
such that $\mu=0,1$, although all the results of this section can
be easily generalized to 3+1D case. The geodesic equations~(\ref{geodesicEq})
become
\begin{equation}
\begin{gathered}\ddot{\gamma}_{0}-\ell e^{-2\ell\left(1-\lambda\right)\gamma_{0}}\left(\dot{\gamma}_{1}+\lambda\ell\dot{\gamma}_{0}\gamma_{1}\right)^{2}=0,\\
\ddot{\gamma}_{1}-2\left(1-\lambda\right)\ell\dot{\gamma}_{0}\dot{\gamma}_{1}-\left(2\lambda-\lambda^{2}\right)\ell^{2}\dot{\gamma}_{0}^{2}\gamma_{1}+\lambda\ell^{2}\gamma_{1}e^{-2\ell\left(1-\lambda\right)\gamma_{0}}\left(\dot{\gamma}_{1}+\lambda\ell\dot{\gamma}_{0}\gamma_{1}\right)^{2}=0.
\end{gathered}
\end{equation}
 The solutions are
\begin{equation}
\begin{gathered}\gamma_{0}\left(s\right)=-\frac{1}{\ell}\ln\left(\frac{\alpha}{\beta}\sinh\left(\ell\beta s+\theta\right)\right),\\
\gamma_{1}\left(s\right)=\left(\frac{\beta}{\ell\alpha}\coth\left(\ell\beta s+\theta\right)+\delta\right)\left(\frac{\alpha}{\beta}\sinh\left(\ell\beta s+\theta\right)\right)^{\lambda}
\end{gathered}
\label{gammaSOL}
\end{equation}
where $\alpha,\beta,\delta,\theta$ are some constants to be determined
by the initial conditions $\gamma_{\mu}\left(0\right)=0$, $\gamma_{\mu}\left(1\right)=p_{\mu}$.
Substituting (\ref{gammaSOL}) in Eq. (\ref{metric}) one finds
\begin{equation}
g^{\mu\nu}\left(\gamma\left(s\right)\right)\dot{\gamma}_{\mu}\left(s\right)\dot{\gamma}_{\nu}\left(s\right)=\beta^{2}=\text{const}.,
\end{equation}
so that, from (\ref{disp-rel-diff}), $m=\beta$. To determine the
value of $\beta$ consider first that, imposing $\gamma_{\mu}\left(0\right)=0$
one gets the relations
\begin{equation}
\alpha=\frac{\beta}{\sinh\left(\theta\right)},\qquad\delta=-\frac{\beta}{\ell\alpha}\coth\left(\theta\right).
\end{equation}
Substituting these in Eqs. (\ref{gammaSOL}) and imposing $\gamma_{\mu}\left(1\right)=p_{\mu}$,
one finds
\begin{equation}
\begin{gathered}p_{0}=-\frac{1}{\ell}\ln\left(\frac{\sinh\left(\ell\beta+\theta\right)}{\sinh\left(\theta\right)}\right),\\
p_{1}e^{\lambda\ell p_{0}}=\frac{1}{\ell}\left(\coth\left(\ell\beta s+\theta\right)\sinh\left(\theta\right)-\cosh\left(\theta\right)\right)
\end{gathered}
\end{equation}
These relations can be inverted and solved for $\beta$ to give the
on-shell relation
\begin{equation}
\mu^2 = \frac{2}{\ell^2}\left( \cosh\left(\ell m\right) - 1  \right)
= {\cal C}_\lambda(p)
= \left(\frac{2}{\ell}\right)^2 \sinh^{2}\left(\frac{\ell p_{0}}{2}\right)-p_{1}^{2}e^{-\ell\left(1-2\lambda\right)p_{0}},
\label{onshellAllOrders1+1D}
\end{equation}
where we defined an ``effective mass'' $\mu$ (in the limit $\ell\rightarrow 0$, $\mu \rightarrow m$).
One can show, by a similar derivation, that the (3+1)D version of (\ref{onshellAllOrders1+1D}) is
\begin{equation}
\mu^2
= {\cal C}_\lambda(p)
= \left(\frac{2}{\ell}\right)^2 \sinh^{2}\left(\frac{\ell p_{0}}{2}\right)-\vec{p}^{2}e^{-\ell\left(1-2\lambda\right)p_{0}} .
\label{onshellAllOrdersAppendix}
\end{equation}

\section{Translations in terms of Poisson brackets}
\label{app:translations}

\subsection{Generic case}
\label{app:translationsGeneric}

Considering spacetime coordinates $\chi^\mu$, having canonical Poisson brackets with the momenta $p_\nu$ ($\left\lbrace p_\mu,\chi^\nu  \right\rbrace = \delta^\nu_\mu$), and spacetime coordinates $x^\mu$ related to the former by some vector field\footnote{We can refer to it as a momentum-space tetrad, but we are not assuming that it project into a tangent flat Minkowski space.} $E_\mu^\nu(p)$, the Poisson brackets between $x^\mu$ and $p_\mu$ follow:
\begin{equation}
x^\mu = E_\nu^\mu (p) \chi^\nu , \qquad \Longrightarrow \qquad
\left\lbrace p_\mu,x^\nu \right\rbrace = E_\mu^\nu (p) .
\label{tetrads}
\end{equation}
Defining the inverse vector $\bar{E}^\nu_\mu: \bar{E}^\nu_\rho E_\mu^\rho = \delta^\mu_b$, the kinetic term contributing to the free particle Lagrangian is then
\begin{equation}
{\cal L}_{kin} = x^\nu \bar{E}_\nu^\mu \dot{p}_\mu.
\label{kineticTermTetrad}
\end{equation}

In the relative locality action the free Lagrangian of an outgoing (incoming) particle $p_\mu$ is integrated in the interval $s \in [-\infty , s_i]$  ($s\in [ s_i, \infty ]$), so that the boundary terms generated by their variation respect to $p_\mu$ couples to the interaction boundary term as
\begin{equation}
\frac{\partial{\cal L}}{\partial\dot{p}_{\mu}} (s_i) = \pm \zeta_{[i]}^\nu \frac{\partial {\cal K}^{[i]}_\nu}{\partial p_\mu} ,
\end{equation}
the $+$ ($-$) sign standing for outgoing (incoming) particle.
The only contribution to the l.h.s. of the last equation comes from the kinetic term, so that from (\ref{tetrads}) and (\ref{kineticTermTetrad}) it follows
\begin{equation}
x^\sigma (s_i)= \pm \zeta_{[i]}^\nu \frac{\partial {\cal K}^{[i]}_\nu}{\partial p_\mu} E^\sigma_\mu (p)
= \pm \zeta_{[i]}^\nu \frac{\partial {\cal K}^{[i]}_\nu}{\partial p_\mu} \left\lbrace p_\mu , x^\sigma  \right\rbrace
= \pm \zeta_{[i]}^\nu \lbrace {\cal K}^{[i]}_\nu , x^\sigma \rbrace .
\label{boundaryCondPB}
\end{equation}
Now, assuming that the translations are implemented as in~\cite{anatomy}, which guarantees translational invariance for causally connected processes, i.e.
\begin{equation}
\zeta_{[i]B}^\mu = \zeta_{[i]A}^\mu - b^\mu,
\label{translationAnatomy}
\end{equation}
it follows that the boundary condition (\ref{boundaryCondPB}) changes as
\begin{equation}
x_B^\mu (s_i)= x_A^\mu (s_i) \pm b^\nu \lbrace {\cal K}^{[i]}_\nu , x^\mu \rbrace .
\end{equation}
The translational invariance~\cite{anatomy} is then implemented extending the translation (\ref{translationEndpoint}) to all the points of the worldline:
\begin{equation}
x_{B}^{\mu} =x_{A}^{\mu} \pm b^{\nu} \{ {\cal K}^{[i]}_{\nu},x^{\mu} \} .
\label{translation}
\end{equation}

\subsection{Time-ordered case}
\label{app:translationsTimeordered}

We consider as explicit example the time-ordered case. The kinetic term is (\ref{kineticTerm}), where in this case $E^\nu_\mu(p) =\xi^\nu_\mu(p)$, defined in~(\ref{killingTranslation}), so that for a particle $(p,x)$, the free Lagrangian is
\begin{equation}
{\cal L}=x^{0}\dot{p}_{0}-x^{j}\left(1-\lambda\right)\ell p_{j}e^{\lambda\ell p_{0}}\dot{p}_{0}+x^{j}e^{\lambda\ell p_{0}}\dot{p}_{j}+{\cal N}\left({\cal C}_{\lambda}\left(p\right)-\mu^{2}\right) .
\end{equation}
Variating the action in terms of $p_\mu$, it generates the boundary term
\begin{equation}
\frac{d}{ds}\left(x^{0}\delta p_{0} + x^{j}e^{\lambda\ell p_{0}}\delta p_{j} - x^{j}\left(1-\lambda\right)\ell p_{j}e^{\lambda\ell p_{0}}\delta p_{0} \right) ,
\label{boundaryKinetic}
\end{equation}
which couples with the interaction boundary term ${\cal K}^{[i]}_\mu(s_i)$ so to give
\begin{equation}
\begin{gathered}
x^{j}\left(s_{i}\right)= \pm \zeta_{[i]}^{\nu}\frac{\partial{\cal K}_{\nu}^{[i]}}{\partial p_{j}}e^{-\lambda\ell p_{0}} , \\
x^{0}\left(s_{i}\right)= \pm \zeta_{[i]}^{\nu}\left(\frac{\partial{\cal K}_{\nu}^{[i]}}{\partial p_{0}}+\left(1-\lambda\right)\frac{\partial{\cal K}_{\nu}^{[i]}}{\partial p_{j}}\ell p_{j}\right) .
\end{gathered}
\label{boundaryCondTO}
\end{equation}
Assuming (\ref{translationAnatomy}), it follows that the boundary conditions (\ref{boundaryCondTO}) transform under translations as
\begin{equation}
\begin{gathered}
x_B^{j}\left(s_{i}\right)= x_A^{j}\left(s_{i}\right) \pm b^{\nu}\frac{\partial{\cal K}_{\nu}^{[i]}}{\partial p_{j}}e^{-\lambda\ell p_{0}} , \\
x_B^{0}\left(s_{i}\right)= x_A^{0}\left(s_{i}\right) \pm b^{\nu}\left(\frac{\partial{\cal K}_{\nu}^{[i]}}{\partial p_{0}}+\left(1-\lambda\right)\frac{\partial{\cal K}_{\nu}^{[i]}}{\partial p_{j}}\ell p_{j}\right) .
\end{gathered}
\end{equation}
Considering now the Poisson brackets (\ref{phaseSpace}) the last expression is nothing but
\begin{equation}
x_{B}^{\mu}\left(s_{i}\right)=x_{A}^{\mu}\left(s_{i}\right)\pm b^{\nu}\left\{ {\cal K}^{[i]}_{\nu},x^{\mu}\right\} .
\label{translationEndpoint}
\end{equation}

\subsection{Behavior under diffeomorphism}
\label{app:translationDiffeo}

Under a diffeomorphism $p_\mu (\tilde{p}_\mu)$ the Lagrangian changes into ${\cal L}^f(\tilde{p}) = {\cal L}(f(\tilde{p})) $, so that it generates the boundary terms
\begin{equation}
\frac{d}{ds}\left(\frac{\partial{\cal L}(f(\tilde{p}))}{\partial\dot{f}_{\nu}}\frac{\partial\dot{f}_{\nu}(\tilde{p})}{\partial\dot{\tilde{p}}_{\mu}}\delta\tilde{p}_{\mu}\right)=\frac{d}{ds}\left(\frac{\partial f_{\nu}(\tilde{p})}{\partial\tilde{p}_{\mu}}\frac{\partial{\cal L}(p)}{\partial\dot{p}_{\nu}}\delta\tilde{p}_{\mu}\right) .
\end{equation}
The interaction boundary term changes into $\tilde{{\cal K}}^f_\mu$, like in~(\ref{boundaryDiffeo}), then the boundary conditions are
\begin{equation}
 \frac{\partial{\cal L}(p)}{\partial\dot{p}_{\nu}} (s_i)= \pm \zeta^{\rho}\frac{\partial\tilde{{\cal K}}^f_{\rho}}{\partial\tilde{p}_{\mu}} \bar{{\cal M}}^\mu_\nu (\tilde{p})
\end{equation}
where again ${\cal M}_\mu^\nu (\tilde{p})= \partial f_\mu (\tilde{p}) / \partial \tilde{p}_\nu $ with inverse $\bar{{\cal M}}_\nu^\mu (\tilde{p})$. 
It follows from (\ref{kineticTermTetrad}) the boundary condition
\begin{equation}
 x^\sigma (s_i)= \pm \zeta^{\rho}\frac{\partial\tilde{{\cal K}}^f_{\rho}}{\partial\tilde{p}_{\mu}} \bar{{\cal M}}^\mu_\nu (\tilde{p}) E^\sigma_\mu(p)
 \label{boundaryCondDiffeo}
\end{equation}
Now notice that under a diffeomorphism the Poisson brackets change as (as expected, $E^\sigma_\mu(p)$ transforms as a vector in the index $\mu$)
\begin{equation}
 E^\sigma_\mu (p)= \{ p_\mu,x^\sigma  \} = \{ f_\mu (\tilde{p}),x^\sigma  \} =  {\cal M}_\mu^\nu (\tilde{p})\{ \tilde{p}_\nu,x^\sigma \},
\end{equation}
so that, substituting the last expression into (\ref{boundaryCondDiffeo}), we get
\begin{equation}
 x^\sigma (s_i)= \pm \zeta^{\nu}\frac{\partial\tilde{{\cal K}}^f_{\nu}}{\partial\tilde{p}_{\mu}} \{ \tilde{p}_\mu , x^\sigma \} = \pm \zeta^{\nu} \{ \tilde{{\cal K}}^f_{\nu}, x^\sigma \} .
 \label{boundaryCondDiffeo-2}
\end{equation}
From the last expression it follows, with the same argument above, that under a diffeomorphism the translated positions are given by the relation
\begin{equation}
 x_B^\sigma = x_A^\sigma \pm b^{\nu}  \{ \tilde{{\cal K}}^f_{\nu}, x^\sigma \} .
 \label{translationDiffeoPB}
\end{equation}

We close this section by noticing that since, from~(\ref{boundaryDiffeo}), $\tilde{{\cal K}}^f_\mu (\tilde{p},\tilde{q},\dots)= {\cal K}_\mu(f(\tilde{p}),f(\tilde{q}),\dots)$, it follows (the obvious result) that
\begin{equation}
 x_B^\sigma = x_A^\sigma \pm b^{\nu}  \{ {\cal K}_{\nu}(p) , x^\sigma \}\Big|_{p=f(\tilde{p})} .
 \label{translationDiffeoPB-repar}
\end{equation}
I.e. the translations in the framework obtained by a (passive) diffeomorphism $p_\mu = f_\mu(\tilde{p})$ are obtained by simply rewriting the translations for the original action in terms of the new momenta variables.


\begin{thebibliography}{50}

\bibitem{DSR}
G.~Amelino-Camelia,
  ``Relativity in space-times with short distance structure governed by an observer independent (Planckian) length scale,''
  Int.\ J.\ Mod.\ Phys.\ D {\bf 11} (2002) 35
  [gr-qc/0012051];
``Testable scenario for relativity with minimum length,''
 Phys.\ Lett.\ B {\bf 510} (2001) 255
 [hep-th/0012238];
 ``Doubly special relativity'',
 Nature \textbf{418} (2002) 34.
gr-qc/0207049.

\bibitem{kowadsr}
J.~Kowalski-Glikman,
``Observer independent quantum of mass,''
Phys.\ Lett.\ A {\bf 286} (2001) 391
hep-th/0102098.

\bibitem{leedsrPRL}
J.~Magueijo and L.~Smolin,
  ``Lorentz invariance with an invariant energy scale,''
  Phys.\ Rev.\ Lett.\  {\bf 88} (2002) 190403
  [hep-th/0112090];
  ``Generalized Lorentz invariance with an invariant energy scale,''
  Phys.\ Rev.\ D {\bf 67} (2003) 044017
  [gr-qc/0207085].

\bibitem{JurekDeSitt}
  J.~Kowalski-Glikman,
  ``De sitter space as an arena for doubly special relativity,''
  Phys.\ Lett.\ B {\bf 547} (2002) 291
  [hep-th/0207279];
  J.~Kowalski-Glikman and S.~Nowak,
  ``Doubly special relativity and de Sitter space,''
  Class.\ Quant.\ Grav.\  {\bf 20} (2003) 4799
  [hep-th/0304101].

\bibitem{FreLiv3Dk}
  L.~Freidel and E.~R.~Livine,
  ``Effective 3-D quantum gravity and non-commutative quantum field theory,''
  Phys.\ Rev.\ Lett.\  {\bf 96} (2006) 221301
   [hep-th/0512113];
  ``Ponzano-Regge model revisited III: Feynman diagrams and effective field theory,''
  Class.\ Quant.\ Grav.\  {\bf 23} (2006) 2021
  [hep-th/0502106].

\bibitem{JurekFrekfield1e2}
L.~Freidel, J.~Kowalski-Glikman and
S.~Nowak, ``From noncommutative kappa-Minkowski to Minkowski space-time,''
Phys.\ Lett.\ B \textbf{648} (2007) 70 [hep-th/0612170];
 ``Field theory on kappa-Minkowski space revisited: Noether charges
and breaking of Lorentz symmetry,'' Int.\ J.\ Mod.\ Phys.\ A
\textbf{23} (2008) 2687 [arXiv:0706.3658 [hep-th]].


\bibitem{ArzanoQfieldCurved}
  M.~Arzano,
  ``Anatomy of a deformed symmetry: Field quantization on curved momentum space,''
  Phys.\ Rev.\ D {\bf 83} (2011) 025025
  [arXiv:1009.1097 [hep-th]].

\bibitem{RelLocPrinciple}
  G.~Amelino-Camelia, L.~Freidel, J.~Kowalski-Glikman and L.~Smolin,
  ``The principle of relative locality,''
  Phys.\ Rev.\ D {\bf 84} (2011) 084010
  [arXiv:1101.0931 [hep-th]];
  ``Relative locality: A deepening of the relativity principle,''
  Gen.\ Rel.\ Grav.\  {\bf 43} (2011) 2547
  [arXiv:1106.0313 [hep-th]].

\bibitem{JurekReview}
  J.~Kowalski-Glikman,
  ``Living in Curved Momentum Space,''
  Int.\ J.\ Mod.\ Phys.\ A {\bf 28} (2013) 1330014
  [arXiv:1303.0195 [hep-th]];
  J.~Kowalski-Glikman,
  ``Curved Momentum Space and Relative Locality,''
  Int.\ J.\ Geom.\ Meth.\ Mod.\ Phys.\  {\bf 09} (2012) 1261008
  [arXiv:1205.1304 [hep-th]].

\bibitem{CarmonaRelLoc}
  J.~M.~Carmona, J.~L.~Cortes, D.~Mazon and F.~Mercati,
  ``About Locality and the Relativity Principle Beyond Special Relativity,''
  Phys.\ Rev.\ D {\bf 84} (2011) 085010
  [arXiv:1107.0939 [hep-th]].

\bibitem{anatomy}
  G.~Amelino-Camelia, M.~Arzano, J.~Kowalski-Glikman, G.~Rosati and G.~Trevisan,
  ``Relative-locality distant observers and the phenomenology of momentum-space geometry,''
  Class.\ Quant.\ Grav.\  {\bf 29} (2012) 075007
  [arXiv:1107.1724 [hep-th]].

\bibitem{FlaGiukRelLoc}
  G.~Gubitosi and F.~Mercati,
  ``Relative Locality in $\kappa$-Poincar\'e,''
  Class.\ Quant.\ Grav.\  {\bf 30} (2013) 145002
  [arXiv:1106.5710 [gr-qc]].

\bibitem{FreidelFieldCurved}
  L.~Freidel and T.~Rempel,
  ``Scalar Field Theory in Curved Momentum Space,''
  arXiv:1312.3674 [hep-th].

\bibitem{GirelliLivineSnyder}F.~Girelli and E.~R.~Livine, ``Scalar
field theory in Snyder space-time: Alternatives,'' JHEP \textbf{1103}
(2011) 132 {[}arXiv:1004.0621 {[}hep-th{]}{]}.

\bibitem{FreidelSnyderRelLoc}
  A.~Banburski and L.~Freidel,
  ``Snyder Momentum Space in Relative Locality,''
  Phys.\ Rev.\ D {\bf 90} (2014) no.7,  076010
  [arXiv:1308.0300 [gr-qc]].

\bibitem{MigSamSnyderRL}
  S.~Mignemi and A.~Samsarov,
  ``Relative-locality effects in Snyder spacetime,''
  Phys.\ Lett.\ A {\bf 381} (2017) 1655
  [arXiv:1610.09692 [hep-th]].

\bibitem{QGphen}
D.~Mattingly,
``Modern tests of Lorentz invariance,''
Living Rev.\ Rel.\  {\bf 8} (2005) 5
[gr-qc/0502097];
S. Liberati, L. Maccione, \lq\lq
Quantum Gravity phenomenology: achievements and challenges\rq\rq
, J.Phys.Conf.Ser. 314 (2011) 012007, arXiv:1105.6234 {[}astro-ph.HE{]};
G. Amelino-Camelia, \lq\lq Quantum gravity phenomenology\rq\rq,
Living Rev.Rel. 16 (2013) 5, arXiv:0806.0339 {[}gr-qc{]}


  \bibitem{BornReciprocity}M.~Born, ``A Suggestion for Unifying Quantum
Theory and Relativity,'' Proc. R. Soc. A \textbf{165} (1938) 291.

\bibitem{snyder}H.~S.~Snyder, ``Quantized Space-Time,'' Phys.
Rev. \textbf{71} (1947) 38.

\bibitem{MajidFoundation}S.~Majid, ``Foundations of Quantum Groups,''
(Cambridge University Press, Cambridge, England, 1995).


\bibitem{whataboutbob}
G.~Amelino-Camelia, M.~Matassa, F.~Mercati and G.~Rosati,
``Taming nonlocality in theories with deformed Poincare symmetry,''
Phys. Rev. Lett. {\bf 106} (2011) 071301,
arXiv:1006.2126.

\bibitem{kappabob}
G.~Amelino-Camelia, N.~Loret and G.~Rosati,
``Speed of particles and a relativity of locality in $\kappa$-Minkowski quantum spacetime.,''
Phys. Lett. {\bf B 700} (2011) 150,
arXiv:1102.4637.


\bibitem{GACboosts}
G.~Amelino-Camelia,
  ``On the fate of Lorentz symmetry in relative-locality momentum spaces,''
  Phys.\ Rev.\ D {\bf 85} (2012) 084034
  [arXiv:1110.5081 [hep-th]].

\bibitem{CarmonaBeySpecRel}
  J.~M.~Carmona, J.~L.~Cortes and F.~Mercati,
  ``Relativistic kinematics beyond Special Relativity,''
  Phys.\ Rev.\ D {\bf 86} (2012) 084032
  [arXiv:1206.5961 [hep-th]];
  J.~M.~Carmona, J.~L.~Cortes and J.~J.~Relancio,
  ``Beyond Special Relativity at second order,''
  Phys.\ Rev.\ D {\bf 94} (2016) no.8,  084008
  [arXiv:1609.01347 [hep-th]].

\bibitem{palmisano}
  G.~Amelino-Camelia, G.~Gubitosi and G.~Palmisano,
  ``Pathways to relativistic curved momentum spaces: de Sitter case study,''
  Int.\ J.\ Mod.\ Phys.\ D {\bf 25} (2016) no.02,  1650027
  [arXiv:1307.7988 [gr-qc]].  

\bibitem{CarmonaRelancioComposition}
  J.~M.~Carmona, J.~L.~Cortés and J.~J.~Relancio,
  ``Relativistic deformed kinematics from momentum space geometry,''
  arXiv:1907.12298 [hep-th].
  
\bibitem{spinning}
  G.~Amelino-Camelia, M.~Arzano, S.~Bianco and R.~J.~Buonocore,
  ``The DSR-deformed relativistic symmetries and the relative locality of 3D quantum gravity,''
  Class.\ Quant.\ Grav.\  {\bf 30} (2013) 065012
  [arXiv:1210.7834 [hep-th]].

\bibitem{causality}
  G.~Amelino-Camelia, S.~Bianco, F.~Brighenti and R.~J.~Buonocore,
  ``Causality and momentum conservation from relative locality,''
  Phys.\ Rev.\ D {\bf 91} (2015) no.8,  084045
  [arXiv:1401.7160 [gr-qc]].

\bibitem{MigRosSnyderRelLoc}
S.~Mignemi and G.~Rosati,
  ``Relative-locality phenomenology on Snyder spacetime,''
  Class.\ Quant.\ Grav.\  {\bf 35} (2018) no.14,  145006
  [arXiv:1803.02134 [gr-qc]].

\bibitem{multipart}
  J.~Kowalski-Glikman and G.~Rosati,
  ``Multi-particle systems in $\kappa$-Poincar\'e inspired by 2+1D gravity,''
  Phys.\ Rev.\ D {\bf 91} (2015) no.8,  084061
  [arXiv:1412.0493 [hep-th]].

\bibitem{CarmonaLocal}
  J.~M.~Carmona, J.~L.~Cortes and J.~J.~Relancio,
  ``Spacetime from locality of interactions in deformations of special relativity: The example of $\kappa$-Poincaré Hopf algebra,''
  Phys.\ Rev.\ D {\bf 97} (2018) no.6,  064025
  [arXiv:1711.08403 [hep-th]].

\bibitem{GiuliaBoosts}
  G.~Gubitosi and S.~Heefer,
  ``Relativistic compatibility of the interacting $\kappa$-Poincar\'e model and implications for the relative locality framework,''
  Phys.\ Rev.\ D {\bf 99} (2019) no.8,  086019
  [arXiv:1903.04593 [gr-qc]].

  
\bibitem{FreSmoGRBrelLoc}
  L.~Freidel and L.~Smolin,
  ``Gamma ray burst delay times probe the geometry of momentum space,''
  arXiv:1103.5626 [hep-th].

\bibitem{meljanac}
S. Meljanac, A. Pachol, A. Samsarov, K. S. Gupta,
``Different realizations of kappa-momentum space and relative-locality effect'',
Phys. Rev. D 87, 125009 (2013),
arXiv:1210.6814 [hep-th].

\bibitem{rovelliBook}
C. Rovelli,
``Quantum gravity'',
Cambridge, UK: Univ. Pr. (2004).

\bibitem{Lukierski}
  J.~Lukierski, H.~Ruegg, A.~Nowicki and V.~N.~Tolstoi,
``Q deformation of Poincare algebra,''
  Phys.\ Lett.\ B {\bf 264} (1991) 331;
  J.~Lukierski, A.~Nowicki and H.~Ruegg,
  ``Real forms of complex quantum anti-De Sitter algebra U-q(Sp(4:C)) and their contraction schemes,''
  Phys.\ Lett.\ B {\bf 271} (1991) 321
  [hep-th/9108018];
 ``New quantum Poincare algebra and k deformed field theory,''
 Phys.\ Lett.\ B {\bf 293} (1992) 344.

\bibitem{MajidRuegg}
  S.~Majid and H.~Ruegg,
  ``Bicrossproduct structure of kappa Poincare group and noncommutative geometry,''
  Phys.\ Lett.\ B {\bf 334} (1994) 348
  [hep-th/9405107].

\bibitem{gacMajid} G. Amelino-Camelia and S. Majid,
``Waves on noncommutative spacetime and gamma ray bursts'',
 Int. J. Mod. Phys. A {\bf 15}, 4301 (2000),
hep-th/9907110.

 
\bibitem{flaviotalk} F.~Mercati, talk given at Perimeter Institute
(available online at http://pirsa.org/11110138/)

\bibitem{CarmonaGoldenRule}
J.~M.~Carmona, J.~L.~Cortes and J.~J.~Relancio,
``Beyond Special Relativity at second order,''
Phys.\ Rev.\ D {\bf 94} (2016) no.8,  084008
[arXiv:1609.01347 [hep-th]].

\bibitem{MajidOecklkFuorier} S.~Majid and R.~Oeckl, ``Twisting
of quantum differentials and the Planck scale Hopf algebra,'' Commun.\ Math.\ Phys.\ \textbf{205}
(1999) 617 {[}math/9811054{]}.

\bibitem{LukKosMaskfield}P.~Kosinski, J.~Lukierski and P.~Maslanka,
``Local D = 4 field theory on kappa deformed Minkowski space,''
Phys.\ Rev.\ D \textbf{62} (2000) 025004 {[}hep-th/9902037{]}.

\bibitem{GACagostinikfield2004} A.~Agostini, G.~Amelino-Camelia
and F.~D'Andrea, ``Hopf algebra description of noncommutative space-time
symmetries,'' Int.\ J.\ Mod.\ Phys.\ A \textbf{19} (2004) 5187
{[}hep-th/0306013{]}.

\bibitem{MeljakBasis}
  S.~Meljanac, A.~Samsarov, M.~Stojic and K.~S.~Gupta,
  ``Kappa-Minkowski space-time and the star product realizations,''
  Eur.\ Phys.\ J.\ C {\bf 53} (2008) 295
  [arXiv:0705.2471 [hep-th]];
S.~Meljanac and S.~Kresic-Juric,
  ``Differential structure on kappa-Minkowski space, and kappa-Poincare algebra,''
  Int.\ J.\ Mod.\ Phys.\ A {\bf 26} (2011) 3385
  [arXiv:1004.4647 [math-ph]];

\bibitem{BoroPachokBasis}
  A.~Borowiec and A.~Pachol,
  ``$\kappa$-Minkowski spacetimes and DSR algebras: Fresh look and old problems,''
  SIGMA {\bf 6} (2010) 086
  [arXiv:1005.4429 [math-ph]].

\bibitem{Kirillov}
A. A. Kirillov, “Lectures on the Orbit Method”, Graduate Studies in Mathematics Vol.64, American Mathematical Society
(2004).

\bibitem{GACMavroGRBs}
  G.~Amelino-Camelia, J.~R.~Ellis, N.~E.~Mavromatos, D.~V.~Nanopoulos and S.~Sarkar,
  ``Tests of quantum gravity from observations of gamma-ray bursts,''
  Nature {\bf 393} (1998) 763
  [astro-ph/9712103].

\bibitem{IceCubeNature}
  G.~Amelino-Camelia, G.~D'Amico, G.~Rosati and N.~Loret,
  ``In-vacuo-dispersion features for GRB neutrinos and photons,''
  Nat.\ Astron.\  {\bf 1} (2017) 0139
  [arXiv:1612.02765 [astro-ph.HE]].
  
  
\end{thebibliography}
\end{document}